\documentclass[11pt,a4paper]{elsarticle}
\usepackage{amsmath,amsfonts,amssymb}
\usepackage{mathtools}
\usepackage{braket}
\usepackage{multirow}
\newcommand{\be}{\begin{equation}}
\newcommand{\ee}{\end{equation}}
\newcommand{\ba}{\begin{eqnarray}}
\newcommand{\ea}{\end{eqnarray}}
\def\bs{\begin{subequations}}
\def\es{\end{subequations}}

\def\cD{{\cal D}}

\def\cF{{\cal F}}
\def\cG{{\cal G}}

\def\cL{{\cal L}}
\def\cM{{\cal M}}

\def\cO{{\cal O}}

\def\cR{{\cal R}}

\def\cZ{{\cal Z}}

\def\p{\partial}

\def\x{q}
\newcommand{\m}{\mu}

\newcommand{\Rb}{\bar{R}}

\newcommand{\bi}{\begin{itemize}}
\newcommand{\ei}{\end{itemize}}

\newcommand{\gb}{\bar{g}}
\newcommand{\eq}[1]{\begin{align} #1 \end{align}}
\newcommand{\spliteq}[1]{\begin{align}\begin{split}\begin{aligned} #1 \end{aligned}\end{split}\end{align}}

\newcommand{\del}{\partial}
\newcommand{\E}{\mathrm{e}}
\newcommand{\Tr}{\operatorname{Tr}}
\newcommand{\diff}{\mathrm{d}}

\renewcommand{\i}{\mathrm{i}}

\newcommand{\hT}{h^{\rm T}}

\begin{document}

\title{RG flows of  Quantum Einstein Gravity \\ in the linear-geometric approximation}

\author[mz]{Maximilian Demmel}
\ead{demmel@thep.physik.uni-mainz.de}
\author[nj]{Frank Saueressig}
\ead{f.saueressig@science.ru.nl}
\author[je]{Omar Zanusso}
\ead{omar.zanusso@uni-jena.de}
\address[mz]{
PRISMA Cluster of Excellence \& Institute of Physics (THEP), \\
University of Mainz, Staudingerweg 7, D-55099 Mainz, Germany
}
\address[nj]{
Institute for Mathematics, Astrophysics and Particle Physics (IMAPP),\\
Radboud University Nijmegen, Heyendaalseweg 135, 6525 AJ Nijmegen, The Netherlands
}
\address[je]{
Theoretisch-Physikalisches Institut, Friedrich-Schiller-Universit\"{a}t Jena, \\
Max-Wien-Platz 1, 07743 Jena, Germany
}

%\date{...}

\begin{abstract}
We construct a novel Wetterich-type functional renormalization group equation for gravity which encodes the gravitational degrees of freedom in terms of
gauge-invariant fluctuation fields. Applying a linear-geometric approximation
the structure of the new flow equation is considerably simpler than the standard Quantum Einstein Gravity construction since only transverse-traceless and trace part of the metric fluctuations propagate in loops. 
The geometric flow reproduces the phase-diagram of the Einstein-Hilbert truncation including the non-Gaussian fixed point essential for Asymptotic Safety.
Extending the analysis to the polynomial $f(R)$-approximation establishes that this 
fixed point comes with similar properties as the one found in metric Quantum Einstein Gravity; in particular it possesses three UV-relevant directions and is stable with respect to deformations of the regulator functions by endomorphisms. 
%In a companion paper we will establish that our flow equation also admits complete fixed functions $f_*(R)$, indicating that the fixed point identified here remains robust when an infinite number of coupling constants is included.
\end{abstract}

% 02.30.Cj	Measure and integration
%	02.40.Gh	Noncommutative geometry
% 02.50.-r	Probability theory, stochastic processes, and statistics
% 04.60.-m	Quantum gravity
% 05.45.Df	Fractals
% 11.10.Kk	Field theories in dimensions other than four
% 11.10.Lm	Nonlinear or nonlocal theories and models
% 11.10.Nx	Noncommutative field theory

%%\pacs{05.45.Df,11.10.Kk,11.10.Lm,04.60.-m,02.50.-r,02.30.Cj,02.40.Gh,11.10.Nx}
%\preprint{MZ-TH/12-34}

%\begin{document}
\maketitle

%%%%%%%%%%%%%%%%%%%%%%%%%%%%%%%%%%%%%%%%%%%%%%%%%%%%%%%%%%%%%%%%%%%%%%%%%%%%%%%%%%%%%%%%%%%%%%%%%%%%%%%%%%%%%%%%%%%%%%%%%%%%%%%%%%%%%%%%%%%%%%%%%%%%%%%%%%%%%%%%%%%%%%%%%%%%%%%%%%%%%%%%%%%%%%%%%%%%%%%%%%%%%%%%%%%%%%%%%%%%%%%%%%%%%%%%%%%%

%--------------------------------------------
\section{Introduction}
%--------------------------------------------
\subsection{Asymptotic safety: a primer}
%--------------------------------------------
Asymptotic Safety \cite{Weinberg:1980gg,Weinproc1,Weinproc2}
provides a natural mechanism
to define a consistent and predictive quantum theory of gravity
within the framework of quantum field theory \cite{Niedermaier:2006wt,Reuter:2007rv,robrev,Reuter:2012id}.
The proposal is conservative in the sense that the gravitational degrees of freedom are carried
by the spacetime metric and invariance under coordinate transformations is retained as a symmetry principle.
The essence of the construction is a non-Gaussian fixed point (NGFP) of the 
 gravitational renormalization group (RG) flow,
which controls the gravitational interactions at high energies. Ideally,
this NGFP should come with a finite number of relevant directions in order to ensure predictivity of the construction.
RG trajectories which are dragged into the NGFP possess a well-defined
UV limit since the dimensionless couplings remain finite
and scattering amplitudes are save from unphysical UV divergences.

Initially proposed by Weinberg, the systematic investigation of Asymptotic Safety started with the
advent of the functional renormalization group equation (FRGE) for the gravitational
 effective average action $\Gamma_k$  
\cite{Reuter:1996cp}
\be \label{the_FRGE}
\p_t \Gamma_k[\Phi, \bar{\Phi}] = \frac{1}{2}  {\rm STr}  \left[ \left( \Gamma_k^{(2)} + \cR_k \right)^{-1} \, \p_t \cR_k \right] \, .
\ee
Here $t \equiv \ln k/k_0$ is the ``renormalization group time''
and $\cR_k$ constitutes an IR regulator, which acts as a mass term for quantum fluctuations of the gravitational field with momenta $p^2 \lesssim k^2$.
The Hessian $\Gamma_k^{(2)}$ denotes the second variation of $\Gamma_k$
with respect to the fluctuation fields $\Phi$ at fixed background fields $\bar{\Phi}$ and is thus a matrix valued inverse propagator in field space.
The trace ${\rm STr}$ contains a sum over loop momenta $p^2$ and internal indices. The regulator dependence in \eqref{the_FRGE}
ensures that the integration over momenta is UV- and IR-finite and ``peaked'' at momenta $p^2 \approx k^2$.
Thus the flow of $\Gamma_k$ is driven by quantum fluctuations at the scale $k^2$ and realizes the
successive integrating out of field modes ``shell-by-shell'' in momentum space as $k$ is lowered.
In fact, taking the limit $k\to 0$ all quantum fluctuations are integrated out and
$\lim_{k \rightarrow 0} \Gamma_k = \Gamma_0$ coincides with the effective action of the theory.

Formally, the FRGE is an exact equation carrying the same information content as the path integral 
from which it is derived in \cite{Wetterich:1992yh} and independently in \cite{Morris:1993qb}.
Moreover, constructing complete solutions $\Gamma_k$ for $k \in [0, \infty[$ is actually equivalent to solving the underlying path integral
or, in other words, to the renormalization of the theory \cite{Litim:2002xm,Codello:2013bra}.
A particular strength of \eqref{the_FRGE} is that it allows to compute approximate solutions for the gravitational RG flow
without 
 resorting to an expansion in a small parameter or presupposing the renormalizability of the theory.
Performing a derivative expansion or vertex expansion of $\Gamma_k$
yields approximate RG flows which are non-perturbative in nature
and whose 
range of validity extends far beyond the Gaussian regime of perturbation theory. These techniques
have played an essential role in establishing confidence in the Asymptotic Safety conjecture.

The first set of evidence supporting the Asymptotic Safety scenario comes from projecting the RG flow entailed by \eqref{the_FRGE}
to a finite number of coupling constants,  
restricting the operators contained in $\Gamma_k$ to a finite subset.
Starting from the seminal works which projected the gravitational effective average action onto the operators included in the
Einstein-Hilbert action \cite{Reuter:1996cp,Dou:1997fg,Souma:1999at},
the gravitational RG flow has been successively projected onto operator subspaces of increasing complexity and field content.
The existence of a NGFP has then been clearly established in the Einstein-Hilbert truncation 
\cite{Lauscher:2001ya,Reuter:2001ag,Nagy:2012rn,Donkin:2012ud} and its extensions including
 the square of the scalar curvature $R^2$ \cite{Lauscher:2001rz,Lauscher:2002sq,Rechenberger:2012pm},
$f(R)$-type polynomial truncations where the effective average action is approximated by  polynomials of the scalar curvature
 \cite{Codello:2007bd,Machado:2007ea,Codello:2008vh,Bonanno:2010bt,Falls:2013bv,Rahmede:2011zz},
the square of the Weyl tensor \cite{Benedetti:2009rx,Benedetti:2009},
 the Gibbon-Hawking-York boundary terms relevant for black holes physics \cite{Becker:2012js},
and in truncations where the quantum effects of the ghost sector are taken into account \cite{Eichhorn:2009ah,Groh:2010ta,Eichhorn:2010tb}.

In particular the polynomial $f(R)$-truncations \cite{Codello:2007bd,Machado:2007ea,Codello:2008vh,Bonanno:2010bt,Rahmede:2011zz,Falls:2013bv}  provide an important indication for the predictivity of Asymptotic Safety: including curvature terms
$R^N$ for $N \ge 3$ in the polynomial ansatz does not unveil new, relevant directions and the number of relevant parameters remains three. In fact, even the initial systematic studies \cite{Lauscher:2002sq,Codello:2007bd,Machado:2007ea} already indicated that the irrelevance of the new directions, as measured by the critical exponents, increases with the order of the polynomial expansion suggesting that power counting might still constitute a good ordering principle at the NGFP. More formal arguments supporting the predictivity of the NGFP have been advocated in \cite{Benedetti:2013jk}.

In contrast to perturbation theory, the FRGE allows to test Asymptotic Safety for various spacetime dimensions
and, in particular, away from both the upper and lower critical dimensions of gravity.
In fact the dimension of spacetime can  be treated as parameter:
the existence of a NGFP can thus be shown for any dimension greater than two \cite{Reuter:2001ag,Fischer:2006at,Ohta:2012vb}
and it is easily seen that the NGFP of Asymptotic Safety 
merges with the Gaussian fixed point at the lower critical dimension $d=2$.
Thus Asymptotic Safety is the simplest
and most effective generalization of Asymptotic Freedom.
Finally, the properties of the NGFP seem to be mostly determined by
the integration of the ``trace-type'' degrees of freedom of the metric,
which are readily visible in the so-called conformally reduced approximation
\cite{Reuter:2008wj,Reuter:2008qx,Reuter:2009kq,Machado:2009ph,Bonanno:2012dg}.
In fact, a recent interpretation by 't Hooft \cite{Hooft:2010nc}
advocates the viewpoint that the conformal degrees of freedom of the metric
should be integrated out first, leaving an effective field theory
for the conformal sector.

More advanced evidence for Asymptotic Safety comes from 
studying expectation values of the fluctuation field (so-called 
``bi-metric'' truncations), signature-dependent effects and anisotropic scaling effects.
Bi-metric truncations explicitly take into account the dependence of $\Gamma_k$ on the fluctuation fields and 
are natural generalizations of the aforementioned ``single-metric'' truncations that provide the first body of evidence.
This program has been initiated in \cite{Manrique:2009uh,Manrique:2010mq} and 
turns out to be crucial for  understanding the background covariance of Asymptotic Safety \cite{Manrique:2010am,Becker:2014qya},
precision computations elucidating the structure of the NGFP \cite{Codello:2013fpa,Christiansen:2012rx,Christiansen:2014raa}
and establishing monotonicity properties of the gravitational RG flow expected from standard RG arguments \cite{Becker:2014pea}.
The dependence of the NGFP on the signature of the metric was first investigated in \cite{Manrique:2011jc},
showing that 
the Asymptotic Safety mechanism is realized independently 
of the signature of the metric \cite{Rechenberger:2012dt}.
Asymptotic Safety has also been showed to play a significant role in  understanding of the phase diagram of the
anisotropic theories of gravity, dubbed Ho\v{r}ava-Lifshitz gravity \cite{Horava:2009uw}.
Ho\v{r}ava-Lifshitz gravity is conjectured to be perturbatively renormalizable and 
the fixed point structure underlying this conjecture as well as its relation to the Asymptotic Safety proposal 
has recently been clarified \cite{D'Odorico:2014iha}.

Progress has also been made in achieving a deeper mathematical and physical understanding
of the Asymptotic Safety mechanism. 
On the one hand, a computer based algorithm for evaluating the derivative expansion
of the gravitational FRGE 
was proposed in \cite{Benedetti:2010nr}
and further developed in \cite{Groh:2011vn,Groh:2011dw}, showing that the expansion is 
 ``computable'' to any order in a strict mathematical sense.
On the other hand, a physical explanation for Asymptotic Safety based on paramagnetic dominance has been advocated in \cite{Nink:2012vd}
which draws a clear and intriguing analogy with the pictorial representation of Asymptotic Freedom
in which charges are screened by virtual pair production.

Paralleling the development of the Asymptotic Safety program based on the FRGE, similarly
encouraging results have been obtained from Monte Carlo simulations using 
dynamical triangulations methods \cite{Loll:1998aj}.
In this case, macroscopic spacetimes are glued together from piecewise linear building blocks (simplices)
and the statistical weight of a configuration is given by the discretized gravitational action.
A particularly successful implementation of this idea are Causal Dynamical Triangulations (CDT) \cite{Ambjorn:2000dv}, reviewed in \cite{Ambjorn:2012jv},
which imprint a causal structure on the triangulations.
Most impressively, the CDT program has established the existence
of a ``classical phase'' where the large-scale properties of the triangulated
geometries resemble those observed in the real world \cite{Ambjorn:2004qm,Ambjorn:2005qt}.
Moreover, there is evidence for a second-order phase transition line
which may allow to take the continuum limit of the underlying lattice theory in a controlled way \cite{Ambjorn:2012ij,Cooperman:2014sca,Cooperman:2014}.
From a RG perspective, the second-order phase transition
may represent the NGFP found by continuum methods, thereby linking CDT and
 the Asymptotic Safety conjecture, also see \cite{Manrique:2008zw,Reuter:2011ah} for a
more detailed discussion of this point.

Based on these findings, the Asymptotic Safety scenario in which gravitational degrees of freedom are captured by fluctuations of the spacetime metric is on firm grounds. This raises the interesting question whether formulating the theory in terms of metric fluctuations is the only possibility to achieve Asymptotic Safety. At the classical level, there are indeed many formulations (metric, first-order formalism, etc.) which give rise to the same dynamics.\footnote{For first studies of the gravitational RG flows employing vielbein-connection variables and the ADM formalism,
see \cite{Harst:2012ni,Harst:2014vca} and \cite{Contillo:2013fua}, respectively.}
It is far from clear, however, that this also holds at the level of the quantum theory.
This question is closely related to the construction of the measure of the gravitational path integral which can crucially influence the content of FRGE. While the FRGE retains its structural form, ${\rm STr}$ and $\Gamma_k^{(2)}$ may  acquire different meanings if the derivation of the FRGE is not based on a field-reparametrization invariant formulation of the path-integral.

%--------------------------------------------
\subsection{Scope of the present work}
%--------------------------------------------
The goal of this paper is the construction of a novel, geometric FRGE based on gauge-invariant fluctuation fields and a specific choice of path-integral measure which are adapted to the fiber-bundle structure of the gravitational configuration space. The structure of the resulting geometric FRGE is significantly simpler than the one underlying non-geometric constructions, since it does not contain contributions from the gauge-fixing, ghost sector and auxiliary traces encoding the Jacobians originating from the transverse-traceless decomposition \cite{York:1973ia} of the fluctuation field. In this sense our construction is more economic than earlier versions of the FRGE based on the Vilkovisky-De Witt formalism \cite{Pawlowski:2003sk,Pawlowski:2005xe,Donkin:2012ud} and avoids the complications associated with evaluating the flow on-shell \cite{Benedetti:2011ct}. In the concrete computations based on the geometric flow equation we truncate the map between the metric fluctuations and gauge-invariant fields at the linear order. This linear-geometric approximation can also be obtained from non-geometric flow equations for a specific choice of gauge-fixing and regulator. Geometrically, this corresponds to an approximation where the Vilkovisky connection on the gauge-bundle is neglected.

 In order to get a first impression of the RG flow encoded in the geometric FRGE,
 we derive the beta functions of the Einstein-Hilbert and $f(R)$-truncation including non-trivial 
endomorphisms in the coarse-graining operator. Based on these beta functions, we 
 identify a NGFP whose properties are remarkably similar to the ones found in non-geometric computations.
In particular, it also comes with three relevant directions 
and the stability coefficients are in good agreement with the ones found in previous studies.
 Our results exhibit only a very mild dependence on the regularization scheme which is mitigated even further when considering observables
such as the critical exponents of the NGFP.
Finally, we exploit the freedom in selecting the coarse-graining scheme and apply the principle of minimum sensitivity (PMS) \cite{Canet:2002gs,Canet:2003qd} in order to
optimize the physics content of the computation by minimizing the scheme-dependence of the observables.
Notably, imposing the PMS conditions significantly improves the convergence of the fixed point properties
within the series of polynomial truncations.

As it will turn out \cite{Demmel:ip}, the field parametrization invariant formulation of the FRGE is closely related to another 
 key challenge currently faced by the Asymptotic Safety program,  the 
extension of the finite-dimensional truncations discussed above to an infinite set of couplings. The latter are implemented by
approximating $\Gamma_k$ by scale-dependent functions. Substituting such an ansatz 
into \eqref{the_FRGE} then yields a non-linear partial differential equation (PDE) which governs
the $k$-dependence of the function contained in the ansatz and the fixed points discussed
above are promoted to $k$-stationary, global solutions of this PDE. The simplest ansatz
of this type approximates the gravitational part of $\Gamma_k$ by a function of
the scalar curvature $R$
\be
\Gamma_k^{\rm grav}[g] = \int d^dx \sqrt{g} f_k(R) \, . 
\ee
This type of functional truncations has been considered in both three \cite{Demmel:2012ub,Demmel:2013myx,Demmel:2014sga}
and four dimensions \cite{Benedetti:2012dx,Benedetti:2013nya}.
The three-dimensional studies demonstrated that the NGFP can consistently be promoted
to a fixed function, and proved, for the first time, that this fixed function admits only a finite number of relevant deformations.
The analogous construction in $d=4$ \cite{Dietz:2012ic,Dietz:2013sba}
has proven to be highly non-trivial, however, and no satisfying fixed function has been found to date.
%These challenges spawned a renewed interest in the construction of flow equations via the background field method,
%which provided deep insights on the flow of the potential in scalar field theories \cite{Bridle:2013sra}.

The remaining parts of the work are organized as follows.
 Sect.\ \ref{Sect.2} contains the derivation of the geometric flow equation,
 discusses its connection to the non-geometric formulations,
 and outlines the approximations required for doing practical computations.
As a first application, the Einstein-Hilbert truncation in the linear-geometric
approximation is investigated in Sect.\ \ref{sect:EH}. 
In Sect.\ \ref{Sect.2.1} we derive a new partial differential equation (PDE) governing
the scale-dependence of $f(R)$-gravity including non-trivial endomorphisms
in the coarse-graining scheme. The fixed point structure entailed by
this PDE is analyzed in Sect.\ \ref{Sect.4} and we conclude with a brief summary and outlook
in Sect.\ \ref{Sect.6}. The technical tools needed for the construction of the beta functions
are summarized in  \ref{app:A}.

%--------------------------------------------
\section{Flow equation for gauge invariant fields}
\label{Sect.2}
%--------------------------------------------
In this section we construct a Wetterich-type functional renormalization group equation for the gravitational effective average action \cite{Reuter:1996cp},
whose flow is solely driven by gauge invariant fields.
The resulting FRGE is independent of the choice of gauge-fixing,
because it invokes a precise cancellation among the un-physical polarization of the metric fluctuations $h_{\mu\nu}$
and the ghost fields themselves.
This leads to significant simplifications of \eqref{the_FRGE} as compared to the past implementations that appeared in the literature,
since the flow does not receive contributions from either ghost or gauge degrees of freedom,
nor from the auxiliary fields that are usually considered to handle the transverse-traceless decomposition of $h_{\mu\nu}$.

%----------------------------------------------------------------------------
\subsection{The flow equation of metric QEG}
%----------------------------------------------------------------------------
The ultimate goal of Quantum Einstein Gravity is to give meaning to a path 
integral over ``all'' metrics $\gamma_{\m\nu}$ suitably weighted by
a bare action $S[\gamma_{\mu\nu}]$.
The bare action is invariant under the general coordinate transformations
\be\label{diffeos}
\delta \gamma_{\mu\nu} = \cL_v \gamma_{\mu\nu} \equiv v^\alpha \, \p_\alpha \gamma_{\mu\nu} + \gamma_{\alpha\nu} \p_\mu v^\alpha + \gamma_{\mu\alpha} \p_\nu v^\alpha \, ,
\ee
which have thus to be suitably factored out from the path integral of the effective action.
Instead of studying the underlying path integral directly, in the Asymptotic Safety program one mainly 
utilizes an effective average action which interpolates smoothly between a UV bare action and the full effective action in the IR.
The renormalization group flow of the effective average action is then governed by a functional RG equation \eqref{the_FRGE}
which can thus be directly used to investigate the theory's properties.
Our construction of the gravitational FRGE is based
on the background field method. In this setting, one splits the quantum metric $\gamma_{\mu\nu}$ into a fixed
(though arbitrary) background metric $\gb_{\mu\nu}$ and corresponding fluctuations $\hat{h}_{\mu\nu}$ 
\be\label{backgroudsplitt}
 \gamma_{\mu\nu} = \gb_{\mu\nu} + \hat{h}_{\mu\nu}.
\ee
The background field formalism then allows to implement a symmetry transformation of the type
\eqref{diffeos} in two different ways. The gauge transformations generalizing \eqref{diffeos} which need
to be gauge-fixed are the {\rm quantum gauge transformations}
\be\label{Qgauge}
\delta_Q \hat{h}_{\mu\nu} = \cL_v \gamma_{\mu\nu} \, , \qquad \delta_Q \gb_{\mu\nu} = 0 \, ,
\ee
which are constructed such that the background metric is left invariant and only the fluctuations transform.
However, a very important feature of the background field method is that it is always possible
to explicitly maintain the
so-called background gauge transformations
\be\label{Bgauge}
\delta_B \hat{h}_{\mu\nu} = \cL_v \hat{h}_{\mu\nu} \, , \qquad \delta_B \gb_{\mu\nu} = \cL_v \gb_{\mu\nu} \, ,
\ee
where the background metric is subject to a background coordinate transformation analogous to \eqref{diffeos}
and every other field, in particular $h_{\mu\nu}$, transforms as a tensor of its corresponding rank.

The fluctuation field is the natural variable of integration in the scale-dependent 
functional integral
\be\label{partsum}
\cZ_k \equiv \int \cD \hat{h}_{\mu\nu} \cD C^\mu \cD \bar{C}_\mu
\exp\Big\{
- \widetilde{S} [\hat{h}, C, \bar{C}; \gb]
%- S[\gb+\hat{h}] - S_{\rm gf}[\hat{h}; \gb] - S_{\rm gh}[\hat{h}, C, \bar{C}; \gb]
- \Delta_kS[\hat{h}, C, \bar{C}; \gb] - S_{\rm source} \Big\} \, ,
\ee
with $\widetilde{S}[\hat{h}, C, \bar{C}; \gb] = S[\gb+\hat{h}] + S_{\rm gf}[\hat{h}; \gb] + S_{\rm gh}[\hat{h}, C, \bar{C}; \gb]$.
In the path-integral we introduced the gauge-fixing term
\be\label{Sgf}
S_{\rm gf} = \frac{1}{2\alpha} \int d^dx \sqrt{\gb} \, \gb^{\mu\nu} \, F_\mu \, F_\nu \, ,  
\ee
which implements a suitable gauge-fixing condition $F_\mu$, and the corresponding ghost term
\be\label{Sghost}
S_{\rm gh}[\hat{h}, C, \bar{C}; \gb] = - \kappa^{-1} \int d^dx \sqrt{\gb} \, \bar{C}_\mu \, \gb^{\mu\nu} \, \frac{\p F_\nu}{\p h_{\alpha\beta}} \, \cL_C(\gb + h) \, ,
\ee
which is obtained from the exponentiation of the Faddeev-Popov determinant.
The path-integral has also been supplemented by source-terms for the fluctuation fields
\be
S_{\rm source} = - \int d^dx \sqrt{\gb} \, \left[ t^{\mu\nu} \, \hat{h}_{\mu\nu} + \bar{\sigma}_\mu C^\mu + \sigma^\mu \, \bar{C}_\mu \right] \, ,
\ee
with which it is possible to construct general expectation values for the corresponding fields.
The key ingredient for constructing the flow equation for the effective average action is the IR regulator $\Delta_k S[\hat{h}, C, \bar{C}; \gb]$
which suppresses fluctuations with momenta smaller than $k^2$ by a $k$-dependent mass term.
In order to explicitly maintain the background symmetry \eqref{Bgauge} at any stage,
it is customary to implement the separation of
high- and low-momentum modes in terms of the eigenvalues of a given covariant operator $\Box$,
which is constructed from the background metric and which encodes the propagation of the fluctuation fields.
In the simplest case (known as Type I cutoff) one chooses the operator to be the background Laplacian $\Box = -\bar{D}^2$,
where the covariant derivative $D_\mu$ is constructed using the Christoffel connection of the background,
however, as we will show later on, other choices are admissible as well.
Generically we choose the IR regulator to be quadratic
 \be
 \Delta_k S = \tfrac{1}{2} \, \int d^dx \sqrt{\gb} \, \hat{\phi} \, \cR_k(\Box) \, \hat{\phi} \, , 
 \ee
where $\hat{\phi} = \{ \hat{h}_{\mu\nu}, C^\mu, \bar{C}_\mu \}$ is the collection of fluctuation fields and $\cR_k$ is matrix valued in field space.

Introducing the $k$-dependent generating functional for the connected Green-functions, $W_k = \ln Z_k$, the vacuum expectation
values of the fluctuation fields are given by the variations of $W_k$ with respect to the corresponding source
\be
h_{\mu\nu} = \frac{1}{\sqrt{\gb}} \, \frac{\delta W_k}{\delta t^{\mu\nu}} \, , \qquad 
\zeta^\mu = \frac{1}{\sqrt{\gb}} \, \frac{\delta W_k}{\delta \bar{\sigma}_\mu} \, , \qquad 
\bar{\zeta}_\mu = \frac{1}{\sqrt{\gb}} \, \frac{\delta W_k}{\delta \sigma^\mu} \, , 
\ee
and will collectively be denoted by $\Phi^i \equiv \left\{ \bar{h}_{\mu\nu}, \zeta^\mu, \bar{\zeta}_\mu\right\}$.
For completeness, we introduce $g_{\mu\nu}$ as the classical analogue of \eqref{backgroudsplitt}
\be
g_{\mu\nu} = \gb_{\mu\nu} + h_{\mu\nu} \,  .
\ee

The effective average action $\Gamma_k[\Phi^i, \bar{\Phi}^i]$ is then defined as the Legendre-transform of $W_k$
up to the subtraction of the IR regulator evaluated on the expectation values
\be
\Gamma_k[\Phi^i, \bar{\Phi}^i] = 
\int d^dx \sqrt{\gb} \left[t^{\mu\nu} \, h_{\mu\nu} + \bar{\sigma}_\mu \zeta^\mu + \sigma^\mu \, \bar{\zeta}_\mu \right] - W_k 
- \Delta_kS[\bar{h}_{\mu\nu}, \zeta^\mu, \bar{\zeta}_\mu; \gb] \, . 
\ee
Following the original derivation \cite{Reuter:1996cp}, one finds that the scale-dependence of the gravitational
effective average action is encoded in the exact functional renormalization group equation \eqref{the_FRGE}.
\iffalse
%
\be\label{FRGE}
\p_t \Gamma_k[\Phi^i, \bar{\Phi}^i] = \frac{1}{2}  {\rm STr}  \left[ \left( \Gamma_k^{(2)} + \cR_k \right)^{-1} \, \p_t \cR_k \right] \, .
\ee
%
Here $t \equiv \ln k/k_0$ is the ``renormalization group time'' and
\fi
Inside \eqref{the_FRGE}, $\Gamma_k^{(2)}$ denotes the second functional derivative of $\Gamma_k$ with respect to the fluctuation fields 
\be\label{fullhessian}
\Gamma_k^{(2)ij}(x,y) \equiv \frac{1}{\sqrt{\gb(x)}} \, \frac{1}{\sqrt{\gb(y)}} \frac{\delta^2 \Gamma_k}{\delta \Phi^i(x) \delta \Phi^j(y)} \, . 
\ee
Eqs.\ \eqref{the_FRGE} and \eqref{fullhessian} conclude our mini-review on the
covariant flow equation for Quantum Einstein Gravity \cite{Reuter:1996cp}.

%----------------------------------------------------------------------------
\subsection{The geometrical flow equation}
%----------------------------------------------------------------------------

From a geometrical perspective the $h_{\mu\nu}$ are coordinates on the configuration
space of the system. For metric QEG this configuration space is given by the fiber bundle Riem($M$)
with the typical fiber being the diffeomorphism group Diff($M$). Physically inequivalent
configurations span the base space Riem($M$)/Diff($M$) of the bundle.

In the previous subsection the sum over physically inequivalent configurations is constructed
by performing the gauge-fixing \eqref{Sgf} and adding the ghost action \eqref{Sghost}. In contrast
to this gauge-fixing procedure, the geometric approach introduces coordinates on Diff($M$) which are 
adjusted to the bundle structure: inequivalent physical configurations are described by their horizontal 
coordinates $\hat{h}^A$ while gauge-equivalent configurations differ by their fiber coordinate $\hat{\varphi}^\alpha$.
By construction the quantum gauge transformations \eqref{Qgauge} act along the fiber
\be
\delta_Q \hat{h}^A = 0 \, , \qquad \delta_Q \hat{\varphi}^\alpha = \hat{\varphi}^{\alpha\prime} \, . 
\ee
Since the action $S[\gamma]$ is diffeomorphism invariant, the transformation entails 
that it must be independent of $\hat{\varphi}^\alpha$: $S[\gamma] = S[h^A; \gb]$ \cite{Vilkovisky:1992pb}.
As a consequence, the analogue of the partition sum \eqref{partsum} may be written as
\be\label{Zgeo}
\cZ_k^{\rm geo} \equiv \int \cD \hat{\varphi}^\alpha \, \int \cD \hat{h}^A 
\exp\Big\{
- S[\hat{h}^A; \gb] - \Delta_kS[\hat{h}^A; \gb] - S_{\rm source} \Big\} \, . 
\ee 
In contrast to the metric construction $\cZ_k^{\rm geo}$ does not include
the gauge-fixing and ghost actions. Moreover, the IR-regulator is
introduced for the gauge-invariant fields $\hat{h}^A$ only. As a consequence
the integration over the fibers becomes trivial and gives rise to an overall
multiplicative factor. Using a suitable definition of $\cD \hat{\varphi}^\alpha$,
\eqref{Zgeo} is invariant with respect to \emph{both} background and quantum gauge
transformations.

Introducing the expectation values $h^A \equiv \langle \hat{h}^A \rangle$ and following the derivation 
of the previous subsection step by step one obtains
the geometric version of the FRGE for Quantum Einstein Gravity 
\eq{\label{geoflow}
\del_t \Gamma_k[h^A; \gb] 
= \frac{1}{2} \Tr\left[ \left( \Gamma_k^{(2)} +  \mathcal{R}_k  \right)^{-1} \, \del_t\mathcal{R}_k \right] \, .
}
While this equation has the same structural form as \eqref{the_FRGE} the crucial difference is the form of the Hessian $\Gamma^{(2)}_k$. While 
the Hessian of the previous construction is with respect to $\Phi^i = \left\{ h_{\mu\nu}, \zeta^\mu, \bar{\zeta}_\mu \right\}$
the $\Gamma^{(2)}_k$ appearing in the geometric flow equation contains the gauge-invariant fields $h^A$ only
\be\label{geohessian}
\Gamma_k^{(2)AB}(x,y) \equiv \frac{1}{\sqrt{\gb(x)}} \, \frac{1}{\sqrt{\gb(y)}} \, \frac{\delta^2 \Gamma_k}{\delta h^A(x) \, \delta h^B(y)} \, . 
\ee

The flow equation \eqref{geoflow} is the central result of this subsection. 
Just like its metric twin, it is an exact equation in the sense
that it contains the same information as the path integral \eqref{Zgeo}.
No approximation has been made in its derivation. Obviously,
eq.\ \eqref{geoflow} is independent of the choice of gauge-fixing
and does not depend on ghost-fields. As its major advantage
it is \emph{invariant under both background and quantum gauge transformations}.

Analogously to the metric flow equation, the solutions of \eqref{Zgeo}
interpolate between the bare action $S[g]$ for $k \rightarrow \infty$
and the standard (off-shell) effective action $\Gamma = \Gamma_{k=0}$ in the IR.
Naturally, all approximation schemes, as, e.g., the derivative expansion or
the vertex expansion, which have been used to construct approximate solutions of \eqref{the_FRGE}
can also be applied to \eqref{Zgeo}. In Sects.\ \ref{sect:EH} and \ref{Sect.4}
we will carry out a first set of checks, verifying that the geometric flow
equation and the metric FRGE leads to similar results.

%----------------------------------------------------------------------------
\subsection{Coordinates on the configuration space}
%----------------------------------------------------------------------------
While the geometrical flow equation \eqref{geoflow} is conceptually nice, its practical usefulness
hinges on being able to construct the map
\be
h_{\mu\nu} \mapsto \left\{ \, h^A \, , \, \varphi^\alpha \, \right\}
\ee
relating the fluctuation field to the coordinates $h^A$ and $\varphi^\alpha$ on
the base space and fiber of the gauge-bundle. In principle, this map can
be constructed by introducing Vilkovisky's connection \cite{Vilkovisky:1984st,Vilkovisky:1992pb} and following 
the construction \cite{DeWitt:1998eq}, also see \cite{tomsparker} for a more
pedagogical exposition and \cite{Pawlowski:2003sk} for a related discussion in the context of the FRGE.

Here we will follow a different path and advocate to construct ``approximately gauge-invariant fields''
via a specific coordinate transformation on the configuration space. Our starting point is the observation
that the quantum gauge transformations \eqref{Qgauge} acting on $h_{\mu\nu}$ is given by
\be\label{Qtrafos}
\begin{split}
\delta_Q \, h_{\mu\nu} = & \, D_\mu \, v_\nu + D_\nu \, v_\mu  \, \\
= & D_\mu \, v_\nu^{\rm T} + D_\nu\, v_\mu^{\rm T}  + 2 D_\mu D_\nu v \, . 
\end{split}
\ee
Here we have performed the decomposition of the coordinate transformation $v_\mu = v_\mu^{\rm T} + D_\mu v$ into its transverse and longitudinal components
{\it with respect to $g_{\mu\nu}$} in the second line. In order to obtain the relation between $h_{\mu\nu}$ and the adapted coordinate
system on the gauge bundle $\left\{h^A, \varphi^\alpha \right\}$
we consider the implicit change of coordinates
 \be
\label{eq:TT}
h_{\mu\nu} = \tilde h_{\mu\nu}^{\rm T} + D_\mu \tilde \xi_\nu + D_\nu  \tilde \xi_\mu + 2 D_\mu D_\nu \tilde \sigma - \tfrac{1}{d} \, g_{\mu\nu} \, \tilde{\chi} \, , 
\ee
where $h_{\mu\nu}[\tilde h_{\mu\nu}^{\rm T}, \tilde \xi_\nu, \tilde \sigma, \tilde \chi; \gb]$ is understood as a functional
of the component fields $\{ \tilde h_{\mu\nu}^{\rm T}, \tilde \xi_\nu, \tilde \sigma, \tilde \chi\}$ and the background metric $\gb$. 
Here $\tilde \chi = \tilde h - 2 D^2 \tilde \sigma$ and the component fields satisfy the constraints
\eq{
	D^\mu \tilde h_{\mu\nu}^{\rm T} =0, \qquad g^{\mu\nu} \tilde h_{\mu\nu}^{\rm T} =0,\qquad D_\mu \tilde \xi^\mu=0,  \qquad g^{\mu\nu} h_{\mu\nu} = \tilde h \, .
}
Comparing \eqref{Qtrafos} with \eqref{eq:TT} one establishes that the component fields inherit the transformation law
\be
\tilde h^{\rm T}_{\mu\nu}  \mapsto \tilde h^{\rm T}_{\mu\nu}  \, , \quad \tilde \xi_\mu \mapsto \tilde \xi_\mu + v^{\rm T}_\mu \, , \quad \tilde \sigma \mapsto \tilde \sigma + v \, , \quad \tilde \chi \mapsto \tilde \chi \, . 
\ee
Thus
\be\label{trivialization}
h^A = \left\{ \, \tilde h^{\rm T}_{\mu\nu} \, , \, \tilde \chi \, \right\} \, , \qquad \varphi^\alpha = \left\{ \, \tilde \xi_\mu \, , \, \tilde \sigma \,  \right\} \, , 
\ee
is the desired trivialization of the configuration space. 

Notably the relation \eqref{eq:TT} is exact and, so far, no approximation has been made: the relation
gives  an \emph{implicit equation} between the metric fluctuations $h_{\mu\nu}$ and the
coordinates $\{h^A, \varphi^\alpha \}$ trivializing the configuration space. Since the r.h.s.\ contains the 
inverse metric $g^{\mu\nu}$ this relation is highly non-linear and non-local in the sense that the relation involves
terms containing an infinite number of derivatives. This reflects the common folklore that there are no local, gauge-invariant
observables in quantum gravity \cite{Torre:1993fq}. Formally, the relation may be solved for the component fields
by applying the projection operators of the transverse-traceless (TT) decomposition \cite{York:1973ia}, replacing $\gb_{\mu\nu}$ with $g_{\mu\nu}$.\footnote{In the case where the TT decomposition is with respect to a fixed background $\gb_{\mu\nu}$ the uniqueness of the
decomposition (up to possible Killing- and conformal Killing vectors of the background) has been established in \cite{York:1973ia}. We are not aware of similarly stringent mathematical results
for the decomposition \eqref{eq:TT}. The fact that \eqref{eq:TT} can be solved by a bootstrap procedure order by order in the fluctuation fields, hints at the possibility that the decomposition is
indeed unique.} These operators involve the inverse of local differential operators, making the non-local nature of the decomposition manifest.

Instead of working with the exact relation \eqref{eq:TT}, we follow the suggestion \cite{Torre:1993fq} 
and use the implicit relation to construct ``approximately gauge-invariant fluctuation fields''.
This strategy actually fits well into the typical approximations used to evaluate
RG flows with the FRGE by performing a derivative expansion and limiting
the power of the fluctuation fields kept track in the flow (``single-metric vs.\ ``bi-metric'' expansions).
Concretely, we will limit ourselves to construct the relation between $h_{\mu\nu}$ and the coordinates \eqref{trivialization}
to linear order. For this purpose, it is useful to perform the standard TT or York-decomposition
of the fluctuation fields with respect to the background metric $\gb_{\mu\nu}$:
 \be
\label{eq:TTdecomposition}
h_{\mu\nu} = h_{\mu\nu}^{\rm T} + \bar{D}_\mu \xi_\nu + \bar{D}_\nu \xi_\mu + 2 \bar{D}_\mu \bar{D}_\nu \sigma - \tfrac{2}{d} \, \bar{g}_{\mu\nu} \, \bar{D}^2 \sigma + \tfrac{1}{d} \, \bar{g}_{\mu\nu} \, h \, ,
\ee
where the component fields are subject to the constraints
\eq{
	\bar{D}^\mu h_{\mu\nu}^{\rm T} =0, \qquad \gb^{\mu\nu} h_{\mu\nu}^{\rm T} =0,\qquad \bar{D}_\mu \xi^\mu=0,  \qquad \gb^{\mu\nu} h_{\mu\nu} = h.
}
Again the terms proportional to $\bar{g}_{\mu\nu}$
can conveniently be combined by introducing the new field
\be
\chi = h - 2 \bar{D}^2 \sigma \, . 
\ee
Notably, the r.h.s.\ of \eqref{eq:TT} and \eqref{eq:TTdecomposition} coincide at linear order in the fluctuation fields.
Thus, at linear order, the coordinates trivializing the configuration space are given
by the component fields of the TT-decomposition:
\be\label{map1}
h^1 = h^{\rm T}_{\mu\nu} + \cO(h^2) \, , \quad h^2 = \chi + \cO(h^2) \, , \quad \varphi^1 = \xi_\mu + \cO(h^2) \, , \quad \varphi^2 = \sigma + \cO(h^2) \, .
\ee
Thus, at this level of approximation the gauge-invariant fields $h^A = \{ h_{\mu\nu}^{\rm T}, \chi \}$
are given by the component fields of the TT-decomposition and the variations in \eqref{geoflow} are 
with respect to the transverse-traceless field and scalar $\chi$ only.

At this stage, the following remark is in order. In order to correctly
make the transition from the gauge-dependent to the gauge-independent coordinates
the knowledge of higher-order terms in the map \eqref{map1} is required. 
A correct single-metric computation (which, by definition, evaluates  \eqref{the_FRGE} at $h_{\mu\nu} = 0$)
requires knowing the quadratic corrections to \eqref{map1}. Consequently, the study of vertex functions including $N$
fluctuation fields (baptized bi-metric computations in \cite{Manrique:2009uh,Manrique:2010mq,Manrique:2010am}) needs the map up to order $N+2$.
The $N$th variation of $h_{\mu\nu}$ with respect to the fields  
$\{ \tilde h^{\rm T}_{\mu\nu}, \tilde \chi, \tilde \xi_\mu, \tilde \sigma \}$
can be constructed systematically from \eqref{eq:TT} by constructing the variations recursively through a bootstrap procedure.
The construction of these terms is beyond the scope of the present paper and we will limit 
our computations to the linear approximation \eqref{map1}. 

% ---------------------------------------------------------------------------
\subsection{The flow equation in Landau-De Witt gauge}
% ---------------------------------------------------------------------------
The flow equation \eqref{geoflow} in combination with the linearized geometric approximation
\eqref{map1} can also be obtained from the standard gauge-fixed covariant flow
equation for QEG \eqref{the_FRGE}, provided that the integration variables 
in the path integral are given by the component fields of the TT-decomposition
and that one evaluates \eqref{the_FRGE} at zero-order in the fluctuation field \cite{grohthesis}.
The first requirements avoids introducing Jacobi-determinants from the
field redefinition while the second approximation allows the explicit cancellation between the
traces containing the gauge- and ghost degrees of freedom.

In order to arrive at \eqref{geoflow} one starts from an ansatz
for the effective average action where the gauge-fixing and ghost terms
retain their classical form
\eq{
	\Gamma_k[\bar{h}, \bar{\zeta}, \zeta; \gb] = \Gamma^{\rm grav}_k[\bar{h}; \gb] + S_{\rm gf}[\bar{h}; \gb] + S_{\rm gh}[\bar{h}, \bar{\zeta}, \zeta; \gb].
}

For the disentanglement of the physical and gauge degrees of freedom
we resort to the so-called minimal TT-decomposition \cite{Benedetti:2010nr},
which refrains from splitting the vector field
into its transverse and longitudinal parts
\be
h_{\mu\nu} = h_{\mu\nu}^{\rm T} + \bar{D}_\mu \rho_\nu + \bar{D}_\nu \rho_\mu - \tfrac{2}{d} \, \bar{g}_{\mu\nu} \, \bar{D}^\alpha \rho_\alpha + \tfrac{1}{d} \, \bar{g}_{\mu\nu} \, h \, .
\ee
The vector field $\rho_\nu$ is related to the component fields appearing in
\eqref{eq:TTdecomposition} by
\be
\rho_\mu = \xi_\mu + \bar{D}_\mu \sigma \, .
\ee 
and thus captures the gauge degrees of freedom at the linear level.
The gauge fixing condition is then chosen 
to be geometrical Landau-De Witt gauge \cite{Mottola:1995sj,Codello:2007bd,Machado:2007ea}

\be
\begin{split}
F_\mu = \bar{D}^\nu \bar{h}_{\mu\nu} - \tfrac{1}{d} \bar{D}_\mu h = \cF[\gb]_\mu{}^\nu \, \rho_\nu \, , 
\end{split}
\ee
with
\be
\cF[\gb]_\mu{}^\nu =  \delta_\mu^\nu \, \bar{D^2}  +  \frac{d-2}{2d} \left(\bar{D}^\nu \bar{D}_\mu  + \bar{D}_\mu \bar{D}^\nu \right)  + \frac{d+2}{2d} \bar{R}_\mu^\nu \, .
\ee
Thus, for this particular gauge choice $F_\mu$ is independent of $h_{\mu\nu}^{\rm T}$ and $h$ and contains $\rho_\nu$
only. Moreover, $\cF$ is hermitian with respect to the ``scalar product'' $\int d^dx \sqrt{\gb}$, so that
\be
S_{\rm gf} = \frac{1}{2\alpha} \, \int d^dx \sqrt{\gb} \, \rho_\mu \, \cG^\mu{}_\nu \, \rho^\nu \, .
\ee
with $\cG = \cF^2$. Setting the background ghosts to zero, the part of the ghost action quadratic in the
fluctuation fields becomes
\be
S_{\rm gh} = - \int d^dx \sqrt{\gb} \, \bar{\zeta}_\mu \, \cM^\mu{}_\nu \, \zeta^\nu   \, . 
\ee
with
\be
\cM^\mu{}_\nu =  \delta_\mu^\nu \, \bar{D^2}  +  \frac{d-2}{2d} \left(\bar{D}^\nu \bar{D}_\mu  + \bar{D}_\mu \bar{D}^\nu \right)  + \frac{d+2}{2d} \bar{R}_\mu^\nu  \, . 
\ee
Thus, for this particular gauge-fixing the ghost operator (at zeroth order in the fluctuation field) agrees with
the gauge fixing operator, $\cF = \cM$.

In the next step, we impose the Landau-De Witt gauge, taking the gauge parameter $\alpha \rightarrow 0$.
In this limit the Hessian $\Gamma^{(2)ij}_k$, given in \eqref{fullhessian}, becomes block-diagonal.
The dynamics of the gravitational sector is governed by the Hessian  $\Gamma^{(2)AB}_k$,
evaluated in the linear geometric approximation, while the contributions of the gauge-degrees
of freedom is determined by the kernel $\cG$ of $S_{\rm gf}$  
\be\label{flowtrunc1}
\begin{split}
\del_t \Gamma_k[h^A, \rho, \bar{\zeta}, \zeta;\gb] 
= & \, \frac{1}{2} \Tr \left( \Gamma^{(2)AB}_k +  \mathcal{R}_k^{AB}  \right)^{-1} \del_t \mathcal{R}_k^{AB} \\ & \, 
+ \frac{1}{2} \Tr \left( \cG +  \mathcal{R}^\cG_k  \right)^{-1} \del_t \mathcal{R}^\cG_k
- \Tr \left( \mathcal{M} + \mathcal{R}^{\rm gh}_k \right)^{-1}\del_t \mathcal{R}^{\rm gh}_k \, . 
\end{split}
\ee
The first line is the r.h.s.\ of the geometric flow equation in the linear geometric approximation. 
The special gauge choice $\cG = \cF^2$ with $\cF = \cM$ implies moreover, that the traces 
in the second line may cancel for a specific choice of regulator. Indeed setting
\be
\mathcal{R}^\cG_k = \left( \mathcal{R}^{\rm gh}_k \right)^2 + \cM \, \mathcal{R}^{\rm gh}_k + \mathcal{R}^{\rm gh}_k \, \cM  \, ,
\ee
and taking into account that $\cM$ is independent of the RG-scale $k$ one has
\be
\frac{1}{2} \Tr \left( \cG +  \mathcal{R}^\cG_k  \right)^{-1} \del_t \mathcal{R}^\cG_k
- \Tr \left( \mathcal{M} + \mathcal{R}^{\rm gh}_k \right)^{-1}\del_t \mathcal{R}^{\rm gh}_k = 0 \, . 
\ee
It can easily be checked that if $\mathcal{R}^{\rm gh}_k$ is a regulator 
then also $\mathcal{R}^\cG_k$ fulfills all properties required of an admissible regulator
function. This is the ``mode-by-mode'' cancellation between
gauge-degrees of freedom and ghost modes used in \cite{Codello:2007bd}.

For the particular choice of regulator \eqref{flowtrunc1} reduces to \eqref{geoflow} subject
to the approximation \eqref{map1}. Thus the geometric flow equation evaluated in the linear geometric approximation
may also be obtained as a particular truncation of the gauge-fixed FRGE \eqref{the_FRGE} with
very specific choices for the gauge-fixing function and regulators. At this stage
we stress that the geometric flow equation \eqref{geoflow} is, in principle,
an exact flow equation for Quantum Einstein Gravity having
the same information content as the standard FRGE but at the same
time being manifestly invariant under background and quantum gauge-transformations.
The RG flows obtained from this flow equation, 
subject to the linear geometric approximation,
will be analyzed in the remaining sections.  
%-----------------------------------------------------------------------------------------------------
%
%
%
%
%-------------------------------------------------------------------------
\section{The Einstein-Hilbert truncation}
\label{sect:EH}
%-------------------------------------------------------------------------
As a first illustration of the RG flows implied by \eqref{geoflow}, we work out
the single-metric Einstein-Hilbert truncation \cite{Reuter:1996cp,Lauscher:2001ya,Reuter:2001ag} 
in the linear-geometric approximation. The corresponding ansatz for the effective average action,
\eq{\label{EH_truncation}
\Gamma_k[h^A; \gb] = \frac{1}{16 \pi G_k} \int\diff^d x  \sqrt{g} \; \left\{ 2 \Lambda_k - R \right\} \, , 
}
contains the scale-dependent Newton's constant $G_k$ and cosmological constant $\Lambda_k$. Note
that in the geometrical formalism $\Gamma_k$ is \emph{not} supplemented by a gauge-fixing and ghost
action and contains $\Gamma^{\rm grav}_k[g]$ only.

Substituting the ansatz \eqref{EH_truncation} into \eqref{geoflow} and
setting the fluctuation field to zero afterwards, the l.h.s.\ of the equation
becomes
\eq{\label{dtEH}
\del_t \Gamma_k[\gb]
=
\int \diff^d x  \sqrt{\gb} \; \left\{ 
2 \, \del_t \left( \tfrac{\Lambda_k}{16 \pi G_k} \right) -  \bar{R} \, \del_t \left( 16 \pi G_k \right)^{-1}
\right\} \, .
}
Thus the scale-dependence of Newton's constant and the cosmological constant is encoded in the coefficients
multiplying the volume term and the interaction term linear in the background Ricci scalar $\bar{R}$. In order to distinguish 
between these terms, it suffices to carry out the computation for the background metric being the one of the $d$-dimensional sphere,
so that the background curvature tensors satisfy \eqref{background}. While the same result can also be
obtained without making a specific choice for $\gb_{\mu\nu}$ \cite{Benedetti:2010nr}, this choice tremendously simplifies
the computation and will be adopted throughout the rest of the paper. 

The beta functions for $G_k$ and $\Lambda_k$ are obtained by evaluating the r.h.s.\ of the flow equation
up to linear order in the background curvature. The first step expands the ansatz \eqref{EH_truncation}
to second order in the fluctuation fields. In the linear-geometric approximation where
$h^A = \{ h_{\mu\nu}^{\rm T}, \chi\}$, $h^\alpha = \{\xi_\mu, \sigma\}$
the quadratic term in this expansion is
\be\label{Gquad}
\begin{split}
\Gamma^{\rm quad}_k [h^A;\gb] = \frac{1}{64 \pi G_k} \int\diff^d x  \sqrt{\gb}\,
\, \bigg\{ & \, h_{\mu\nu}^{\rm T} \left[ \Delta  - 2 \Lambda_k + c_T \bar{R} \right] h^{{\rm T}\mu\nu} \\ & \, 
- \tfrac{(d-2)(d-1)}{2} \chi \left[ \Delta - \tfrac{d}{d-1}\Lambda_k + c_S \, \bar{R} \right] \chi \, \bigg\}
\end{split}
\ee
with
\be
c_T \equiv \frac{d^2-3d+4}{d(d-1)} \, , \qquad c_S \equiv \frac{d-4}{2(d-1)} \, .  
\ee
Thus the Hessian \eqref{geohessian} obtained from \eqref{Gquad} is diagonal in field space.

The next step is the construction of the regulator $\cR_k$. For definiteness, we choose 
the coarse-graining operator $\Box = \Delta$ and implement a Type I regulator scheme \cite{Codello:2008vh}.
In this case $\cR_k$ is determined by the replacement rule $\Delta \mapsto P_k \equiv \Delta + R_k$
with $R_k$ a scalar cutoff function suppressing low-energy fluctuations by a mass-term.
This implicit definition fixes
\be\label{EHcutoff}
\cR_k = \frac{1}{32\pi G_k} \, {\rm diag} \left[ R_k \, {\bf 1} \, , \, - \tfrac{(d-2)(d-1)}{2} \, R_k \, \right] \, .
\ee
The truncated flow equation resulting from the ansatz \eqref{EH_truncation} then reads
\eq{
\label{eq:flowEH}
\del_t \Gamma_k [\gb]
=
\frac{1}{2} {T}_{(2)}
+
\frac{1}{2} {T}_{(0)}
}
with the traces $T_{(s)}$ in the transverse-traceless ($s=2$) and scalar ($s=0$) sector 
given by
\be\label{T1trace}
\begin{split}
{T}_{(2)} = & \, \Tr_{(2)}\left[ \Big(P_k - 2\Lambda_k +  c_T \bar{R} \Big)^{-1} \Big(\del_t R_k - \eta \, R_k\Big) \right] \, , \\
{T}_{(0)} = & \, \Tr_{(0)}\left[ \Big(P_k - \tfrac{d}{d-1}\Lambda_k + c_S \bar{R}\Big)^{-1} \Big( \del_t R_k - \eta \, R_k \Big) \right] \, .
\end{split}
\ee
Here $\eta \equiv \p_t \ln G_k$ denotes the anomalous dimension of Newtons constant.

The evaluation of the traces up to linear order in the background curvature is readily
done by applying the trace technology of \ref{app:A2}. Expanding the trace arguments
in $\bar{R}$, the functions $W(\Delta)$ entering into eq.\ \eqref{mastertrace} are of 
the form
\be
W_m(z; w) \equiv \left( P_k + w \right)^{-m} \, \left( \p_t \, R_k - \eta \, R_k \right) \, . 
\ee
This motivates defining
\eq{\label{Qdef2}
Q^m_n(w) \equiv Q_n\left[ \, W_m(z; w) \, \right] \, . 
}
Applying the expansion \eqref{mastertrace} and retaining the
terms contained in the ansatz \eqref{EH_truncation} only, the traces
\eqref{T1trace} give
\spliteq{\label{traceeval}
T_{(2)}
&=
\frac{1}{(4\pi)^{d/2}}
\int d^dx \sqrt{\gb}
\bigg[
Q^1_{d/2}(-2\Lambda_k)\, b^{(2)}_0
\\
&\qquad
+
\left.
  \left( Q^1_{d/2-1}(-2\Lambda_k)\, b^{(2)}_1 - c_T \, Q^2_{d/2}(-2\Lambda_k)\, b_0^{(2)}
\right) \bar{R}  
\right],
\\
T_{(0)}
&=
\frac{1}{(4\pi)^{d/2}}
\int d^dx \sqrt{\gb}
\bigg[
Q^1_{d/2}\!\left(-\tfrac{d \, \Lambda_k}{d-1}\right)\, b^{(0)}_0
\\
&\qquad
+
\left.
  \left( Q^1_{d/2-1}\!\left(-\tfrac{d \, \Lambda_k}{d-1}\right)\, b^{(0)}_1 - c_S \, Q^2_{d/2}\!\left(-\tfrac{d \, \Lambda_k}{d-1}\right)\, b_0^{(0)}
\right) \bar{R}  
\right] \, ,
}
with the coefficients $b_n^{(s)}$ given in \eqref{eq:EHhkcoeff}.

The beta functions capturing the scale-dependence of $G_k$ and $\Lambda_k$
are then obtained by substituting \eqref{dtEH} and \eqref{traceeval} into \eqref{eq:flowEH}
and equating the coefficients multiplying different powers of the scalar curvature.
The result is most conveniently be expressed in terms of the dimensionless
coupling constants
\be
\lambda_k \equiv  \Lambda_k \, k^{-2} \; , \qquad
g_k \equiv G_k \, k^{d-2} \, ,
\ee
and reads
\be\label{defbeta}
\p_t g_k = \beta_g(g, \lambda) \, , \qquad \p_t \lambda_k = \beta_\lambda(g, \lambda) \,  ,
\ee
with
\be\label{betafcts}
\begin{split}
\beta_g = & \, (d-2+\eta) g \\
\beta_\lambda = & (\eta -2) \lambda
+\frac{g}{(4\pi)^{d/2-1}}
\left[
 (d-2) (d+1)\Phi^1_{d/2} \left(-2 \lambda \right)+ 2 \Phi^1_{d/2} \left(-\tfrac{d \lambda}{d-1} \right)
\right.
\\
&\qquad \qquad \qquad \qquad \qquad
\left.
-\eta  \left(\tfrac{(d-2) (d+1)}{2} \tilde{\Phi }^1_{d/2}\left(-2 \lambda \right)+ \tilde{\Phi
   }^1_{d/2}\left(-\tfrac{d \lambda }{d-1}\right)\right)
\right] \, . 
\end{split}
\ee
The explicit expression for the anomalous dimension of Newton's constant is given by
\be\label{etaresult}
\eta = \frac{g B_1(\lambda)}{1-g B_2(\lambda)} \, , 
\ee
with
\spliteq{
B_1(\lambda)
&=
\frac{1}{(4\pi)^{d/2-1}}
\Bigg[
\tfrac{\left(d^2+3 d+2\right) \left(3 \delta _{d,2}+d-5\right) }{3 (d-1)}\Phi^1_{d/2-1} \left(-2 \lambda \right)
+\tfrac{2}{3} \Phi^1_{d/2-1}\left(-\tfrac{d \lambda }{d-1}\right)
\\
&\qquad
-\tfrac{2 \left(d^4-4 d^3+5 d^2+2 d-8\right)}{d(d-1)} \Phi^2_{d/2} \left(-2 \lambda \right)
-\tfrac{2(d-4) }{d-1}\Phi^2_{d/2} \left(-\tfrac{d \lambda }{d-1}\right)
\Bigg]
\\
B_2(\lambda)
&=
\frac{1}{(4\pi)^{d/2-1}}
\Bigg[
-\tfrac{\left(d^2+3 d+2\right) \left(3 \delta _{d,2}+d-5\right) }{6 (d-1)}\tilde{\Phi }^1_{d/2-1}\!\left(-2 \lambda \right)
-\tfrac{1}{3} \tilde{\Phi}^1_{d/2-1}\!\left(-\tfrac{d \lambda }{d-1}\right)
\\
&\qquad
+\tfrac{\left(d^4-4 d^3+5 d^2+2d-8\right) }{d(d-1)}\tilde{\Phi}^2_{d/2}\left(-2 \lambda \right)
+\tfrac{d-4}{d-1}\tilde{\Phi}^2_{d/2}\left(-\tfrac{d \lambda }{d-1}\right)
\Bigg]
}
In these expressions, the dependence of the beta functions on the regulator is captured
by the dimensionless threshold functions
\be
\begin{split}
\label{eq:thresholds}
\Phi^m_n(w) & \equiv
\frac{1}{\Gamma(n)} \int_0^\infty\diff y\, y^{n-1}\; \frac{\varrho(y) - y \varrho'(y)}{(y+\varrho(y)+w)^m},
\\
\tilde{\Phi}^m_n(w)
&\equiv
\frac{1}{\Gamma(n)} \int_0^\infty\diff y\, y^{n-1}\; \frac{\varrho(y)}{(y+\varrho(y)+w)^m} \, , 
\end{split}
\ee
where $\varrho(y)$ denotes the dimensionless profile function of the cutoff, $R_k = k^2 \varrho(z k^{-2})$. These
threshold functions are related to the $Q$-functionals \eqref{Qdef2} by
\eq{
Q^m_n(w) = k^{2(n-m)+2} \, \left[ 2 \Phi^m_n(w/k^2) - \eta \, \tilde{\Phi}^m_n(w/k^2) \right] \, .
}
The beta functions \eqref{betafcts} together with the expression for the anomalous dimension of Newton's constant
\eqref{etaresult} constitute the main result of this section.

For the remainder of this section we then study the properties of the flow implied by
\eqref{defbeta}. For this purpose we set $d=4$ and chose the optimized regulator \cite{Litim:2001up}
\eq{\label{Rkopt}
\varrho(z) = (1-z)\theta(1-z) \, .
}
For this choice of regulator, the integrals appearing in the threshold functions
can be performed analytically, yielding
\be
\Phi^m_n(w)
=
\frac{1}{\Gamma(n+1)}\frac{1}{(1+w)^m}
\; , \qquad 
\tilde{\Phi}^m_n(w)
=
\frac{1}{\Gamma(n+2)}\frac{1}{(1+w)^m}.
\ee

The most important feature of an RG flow
are its fixed points (FPs) $g_*$ where,
by definition, all beta functions vanish
simultaneously  $\left. \beta_{g_i} \right|_{g= g_*} = 0$.
Given a FP, the dynamics of the RG flow in its vicinity
is captured by the linearized system
\be
\p_t \, g_i = {\bf M}_{ij} \, \left(g_j - g_{j*} \right) \, , 
\ee
governed with the stability matrix
\be\label{defstab}
{\bf M}_{ij} = \left. \frac{\partial \beta_{g_i}}{\partial g_j} \right|_{g = g_*} \, . 
\ee
The critical exponents $\theta_i$, defined as minus the eigenvalues of ${\bf M}$,
are a characteristic feature of the FP, identifying its universality class. In particular
they encode whether a perturbation of the FP theory is UV relevant (Re$(\theta) > 0$),
irrelevant (Re$(\theta) < 0$), or marginal (Re$(\theta) = 0$).

Inspecting the beta functions \eqref{betafcts}, one finds that the system possesses
a so-called Gaussian Fixed Point (GFP)
\be
{\rm GFP:} \qquad g_* = 0 \, , \qquad \lambda_* = 0 \, .
\ee
This point corresponds to a non-interacting theory and the critical exponents of the fixed point
match the mass dimension of the coupling constants. In addition, the beta functions
possess a non-Gaussian Fixed Point (NGFP) with $g_* > 0, \lambda_* > 0$. The position
and characteristic properties of this fixed point obtained from applying the linear-geometric
approximation to the Einstein-Hilbert truncation are summarized in the first line
of Table \ref{tab:EHfp}.
\begin{table}[t]
\begin{center}
\begin{tabular}{|c|c|c|c|c|l|}
\hline
$g_*$ & $\lambda_*$ & $g_* \lambda_*$ & $\theta^\prime$ & $\theta^{\prime\prime}$ & \\ \hline \hline
$0.781$ & $0.203$   & $0.16$ & $2.929$ & $2.965$  & linear geometric flow \\ \hline \hline
$0.272$ & $0.348$   & $0.12$ & $1.547$ & $3.835$ & TT-decomposition \\ \hline
$0.403$ & $0.330$   & $0.13$ & $1.941$ & $3.147$ & harmonic gauge \\ \hline
$1.178$ & $0.25$ & $0.29$ & $1.667$ & $4.308$ & optimized flow \\ \hline
$0.893$ & $0.164$ & $0.146$ & $2.03$ & $2.69$ & geometrical background flow \\ \hline
$1.692$ & $0.144$ & $0.244$ & $1.34$ & $2.61$ & geometrical dynamical flow 
\\ \hline \hline
$2.665$ & $0.415$ & $1.11$ & $1.471$ & $9.304$ & CREH (pot) \\ \hline
$4.650$ & $0.279$ & $1.30$ & $4.0$ & $6.184$ & CREH (kin) \\ \hline
\end{tabular}
\caption{
\label{tab:EHfp} Characteristic features of the NGFP appearing in Einstein-Hilbert truncation. The results
based on the linear geometric approximation introduced in this work are given in the first line. The lower
lines show characteristics of the NGFP obtained from the gauge-fixed Einstein-Hilbert truncation
\cite{Reuter:1996cp,Lauscher:2001ya,Reuter:2001ag,Litim:2003vp,Donkin:2012ud}
and the conformally reduced Einstein-Hilbert (CREH) truncation \cite{Reuter:2008wj} for comparison.
}
\end{center}
\end{table}
For comparison we also include the characteristic properties
of the NGFP found in previous literature studies supplementing
the Einstein-Hilbert action by gauge-fixing and ghost terms.
Notably, the geometric flow recovers all the characteristic 
properties found previously, including the complex
pair of critical exponents 
$\theta_{1,2} = \theta^\prime \pm i \theta^{\prime\prime}$.
Moreover, the universal combination $g_* \lambda_*$ 
essentially agrees with the value found in the full, gauge-fixed 
case. We take these findings as a strong indication
that the properties of the NGFP are largely
governed by the transverse-traceless and trace-sectors
of the flow equation while the gauge-fixing
and ghost contributions play a minor role only.

\begin{figure}[t]
\begin{center}	
\includegraphics[width=0.6\textwidth]{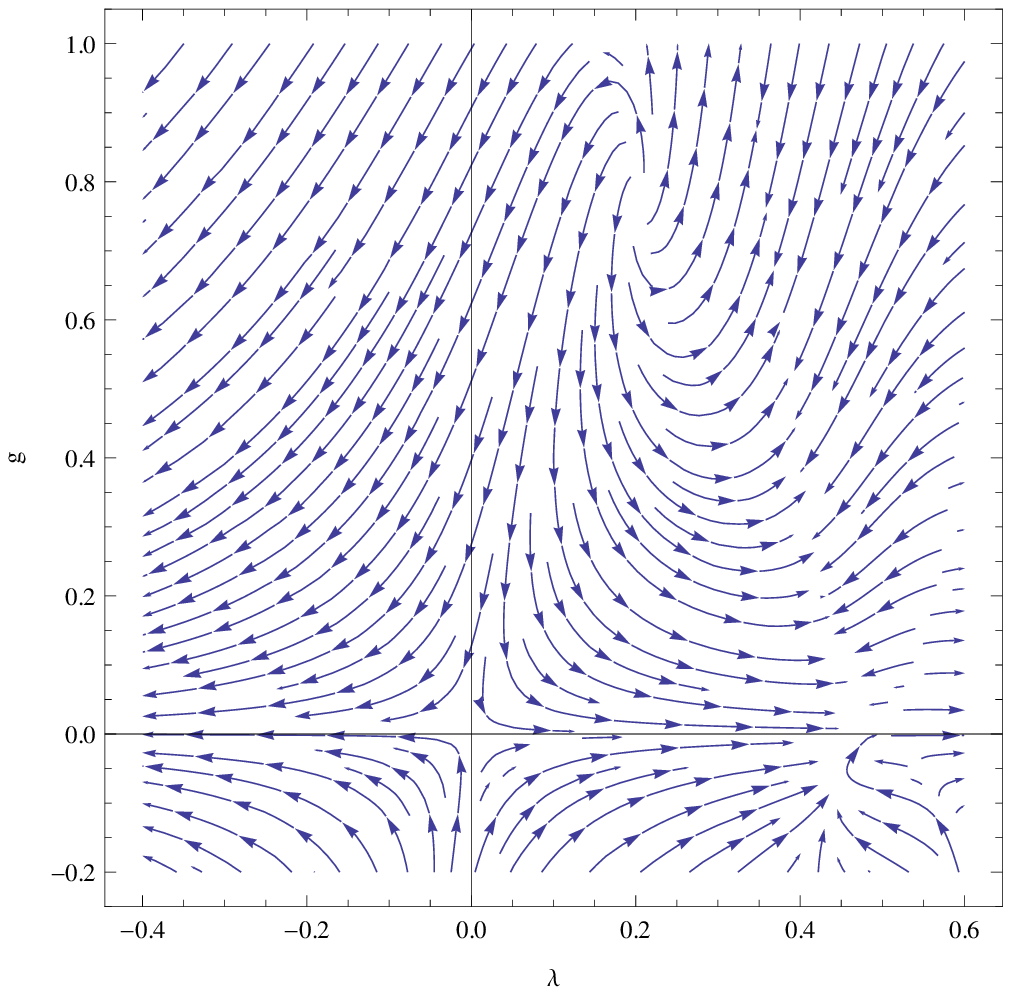}
\caption{
\label{fig:EH_phasediagram}
 The flow of the dimensionless cosmological constant $\lambda_k$ and Newton's coupling $g_k$. The arrows point towards
 the IR, i.e.\ decreasing values of $k$.
}
\end{center}
\end{figure}
In addition to studying the characteristic features of the NGFP appearing in the linearized geometric approximation,
we also construct the phase diagram obtained from integrating the flow \eqref{defbeta}. The flow is shown in
Fig.\ \ref{fig:EH_phasediagram}. For $g_k > 0$ it is governed by the interplay between the NGFP, governing the UV-behavior of the theory, and GFP
controlling its IR-regime. The phase diagram qualitatively displays all the features encountered in previous RG studies based 
on the Einstein-Hilbert truncation \cite{Reuter:2001ag}. The qualitative match of the fixed point characteristics and phase diagrams provides
a strong indication that our new flow equation based on gauge-invariant fields \eqref{geoflow} indeed captures
all essential features of the gravitational RG flow even in the case where the
linear geometric approximation \eqref{map1} is invoked. In particular, 
the results are significantly closer to the gauge-fixed computations than the one
taking into account the contribution of the conformal factor only (CREH).
%-----------------------------------------------------------------------------------------------------
%
%
%
%
%
%
%
%-----------------------------------------------------------------------------------------------------
\section{The $f(R)$-truncation: flow equation}
\label{Sect.2.1}
%-------------------------------------------------------------------------
Upon completing our analysis of the RG flows
obtained from the Einstein-Hilbert truncation in the
linear-geometric approximation, we shall now derive
the partial differential equation (PDE) governing
the RG flow of $f(R)$-gravity. The essentially
new ingredient in the derivation is the inclusion
of non-trivial endomorphisms in the regulator. We
will find that the choice of endomorphism may alter
the singularity structure of the flow equation,
so that the existence of isolated fixed functionals
becomes admissible.

The $f(R)$-truncation approximates the effective average action
by
\eq{\label{fofR_truncation}
	\Gamma_k[h^A;\gb] = \int \diff^d x \sqrt{g} \, f_k(R)\, .
}
Again the ansatz for $\Gamma_k$ depends on the gauge-invariant fluctuation
fields $h^A = \{\hT_{\mu\nu},\chi\}$ only and does not include gauge-fixing and ghost-terms.
The ansatz generalizes \eqref{EH_truncation} by including infinitely many scale-dependent couplings and 
 thus captures significantly more information about Asymptotic Safety, 
for instance concerning the predictive power of the construction.

The construction of the PDE governing 
the scale-dependence of $f_k(R)$ essentially parallels the 
computation of the beta functions for the Einstein-Hilbert truncation.
Substituting the ansatz \eqref{fofR_truncation} into the geometric
flow equation \eqref{geoflow} and subsequently setting the 
fluctuation fields to zero the l.h.s.\ becomes
\be\label{frrhs}
\p_t \Gamma_k = \int \diff^d x \sqrt{\gb} \, \p_t f_k(\Rb) \, . 
\ee
This implies in particular that we can again chose the background metric to be
the one of the $d$-sphere $S^d$, which suffices to project
the exact RG flow on subspace spanned by the
ansatz \eqref{fofR_truncation}. In this background
the part of $\Gamma_k$ quartic
in gauge invariant quantities $h^A = \{\hT_{\mu\nu},\chi\}$
is given by
\be
\begin{split}
\Gamma^{\rm quad}_k[h^A;\gb] 
=
\frac{1}{2} \, \int \diff^d x  \sqrt{\gb} \, & \bigg\{   -\frac{1}{2} 
\hT_{\mu\nu}
\left[
\left( \Delta - \tfrac{2(d-2)}{d(d-1)} \bar{R} \right) f' + f
\right]
{\hT}^{\mu\nu}
\\
&\quad +
\tfrac{d-2}{4d} 
\chi\Big[
\tfrac{4(d-1)^2}{d(d-2)}
%\left( \Delta -\tfrac{\bar{R}}{d-1}\right)^2
\Delta^2_1
f'' 
+\tfrac{2(d-1)}{d}
%\left( \Delta -\tfrac{2\bar{R}}{d-1}\right)
\Delta_2
f'
+
f
\Big]
\chi 
\bigg\} \, ,
\end{split}
\ee
Here the prime denotes a derivative with respect to $\Rb$ and we omitted
the $k$- and $\bar{R}$-dependences of $f_k(\bar{R})$ for notational 
clarity.
In the second line we introduced the abbreviation
$
\Delta_n \equiv \Delta - \tfrac{n\bar{R}}{d-1}
$.

The next step consists in constructing the regulator $\cR_k$. Here
we generalize the computation of the last section by allowing for
a non-trivial spin-dependent endomorphism $E_{(s)}$ in the coarse-graining
operator:
\be
\Box \equiv -\bar{D}^2 + E_{(s)} \, . 
\ee
For practical reasons, we assume that $E_{(s)}$ is covariantly constant with
respect to $\bar{D}_\mu$. Including $E_{(s)}$ brings the advantage that one can shift the value $k_0$ 
where all fluctuation fields are integrated out. In particular
the choice 
\be\label{elev}
E_{(0)} = E_{(2)} + \tfrac{2}{d-1} \, \bar{R}
\ee
 entails that the fluctuations in the scalar ($s=0$) and transverse-traceless $(s=2)$ coming with the lowest eigenvalues  are
 integrated out at the same value $k_0$. Regulators that obey the relation \eqref{elev} thus satisfy the condition of 
equal lowest eigenvalues (ELE). Subsequently, we define the regulator $\mathcal{R}_k$ by the replacement rule $\Box \mapsto P_k(\Box) \equiv \Box + R_k (\Box)$.
This implicit definition entails
\eq{
	\mathcal{R}_k = \operatorname{diag}\left[ \, 
		R_{k}^{(2)} \, \mathbf{1} \, , \, 
		R_{k}^{(0)} \, 
		\right],
}
with
\be
\begin{split}
R_k^{(2)}  & \, = -\tfrac{1}{2} \left(P_k - \Box \right) \, f' \, , \\
R_k^{(0)} & \, = \tfrac{(d-1)^2}{d^2} \left( f''\, \big(P_k^2 - \Box^2 \big) + \big(\tfrac{d-2}{2(d-1)} f' - 2 \left( E_{(0)} + \tfrac{\bar{R}}{d-1} \right) f'' \big) \big(P_k - \Box \big) \right).
\end{split}
\ee

We will now specialize to the case $d=4$. The projected flow equation is then again of the form \eqref{eq:flowEH} with the
transverse-traceless and scalar trace given by
\spliteq{\label{fullequations}
T_{(2)}
&= 6 \, 
\Tr \left[
\frac{
\del_t R_k^{(2)}
}{
(3 E_{(2)} + \bar{R}- 3 P_k ) f' -  f 
}
\right] \, ,
\\
T_{(0)}
&=
16 \, \Tr \left[
\frac{
\del_t R_k^{(0)}
}{
\left(3 E_{(0)} +\bar{R}-3 P_k\right){}^2 f'' - \left(3 E_{(0)} + 2 \bar{R} - 3 P_k\right) f' +2 f 
}
\right] \, .
}
For vanishing endomorphisms these traces coincide with the transverse-traceless
and scalar sectors of the flow equations in the $f(R)$-truncation derived in \cite{Machado:2007ea,Codello:2008vh}.
The crucial difference to these works is the inclusion of an arbitrary endomorphism
in the regulator (Type II cutoff) and the absence of the auxiliary sector, capturing the
Jacobians from the transverse-traceless decomposition of the metric fields.

The operator traces \eqref{fullequations} depend on the coarse-graining operator $\Box$ and
can be evaluated by slightly modifying eq.\ \eqref{meltraf0}
\be\label{eq:Def:traces}
 \Tr \left[ W(\Box) \right]
= 
 \int\limits_0^\infty \diff \sigma\; \widetilde{W}(\sigma) \, \E^{-\sigma E_{(s)}} \, \Tr \left[ \E^{-\sigma\Delta} \right] \, . 
\ee
Here $\widetilde{W}$ denotes the inverse Laplace transform of $W$ and we have used
that the endomorphism $E_{(s)}$ is covariantly constant in order to 
extract it from the operator trace. The factor $\operatorname{T}_{E_{(s)}} \equiv \E^{-\sigma E_{(s)}}$
represents the translation operator on Laplace space. It acts on functions
by a shift of their argument $\operatorname{T}_a f(x) = f(x+a)$.
Combining this feature with the early-time expansion of the heat-kernel on
the sphere \eqref{eq:asymptotic}, the trace can be rewritten
in terms of the Q-functionals defined in \eqref{Qdef}
\eq{
\label{eq:trace_expansion}
\Tr_{(s)} \left[ W(\Box) \right]
=
\frac{1}{(4\pi)^2}\int\diff^4 x \sqrt{\gb}\;\sum_{n\geq 0} Q_{2-n}\left[\operatorname{T}_{E_{(s)}}\!\!W\right] \, b^{(s)}_n \, \bar{R}^n.
}
The coefficients $b^{(s)}_n$ in this expansion can be obtained by summing the eigenvalues of the Laplace-operator
on the sphere and are listed in Table \ref{tab:HKcoeffs} of Appendix \ref{app:A}. For $d=4$ the index
of the Q-functionals is integer. In this case the $Q_m$ are related to the function $W$ via eq.\ \eqref{Qeval1}. 

The last missing ingredient in writing down the PDE governing the scale-dependence of $f_k(R)$
is the specification of the regulator $R_k$. Following the previous section, we will again
adopt the optimized cutoff \eqref{Rkopt}. This choice has the advantage
that only a finite number of terms in the expansion \eqref{eq:trace_expansion} 
actually contribute to the RG flow \cite{Codello:2007bd,Machado:2007ea,Codello:2008vh}. 
This can be seen as follows. For the specific choice of regulator, the arguments of the $Q$-functionals
 appearing in \eqref{eq:trace_expansion} are of the form 
\eq{
\operatorname{T}_{E_{(s)}}\!\!W_s(z)
%= W_s(z+ E_{(s)}) 
= \left[A (z + E_{(s)})^2 + B (z+ E_{(s)}) + C\right] \theta(k^2-E_s - z) \, .
}
The coefficients $A$, $B$ and $C$ depend on the scalar curvature and are independent of $z$. For negative index,
the $Q$-functionals are given by derivatives of this function, evaluated at $z=0$. Owed to the 
polynomial form, only the first two derivatives are contributing to the flow; 
derivatives acting on $\theta(k^2-E_{(s)} - z)$ produce (derivatives of) $\delta$-distributions, $\delta^{(m)}(k^2-E_{(s)})$, which are
outside the truncation subspace. Thus the traces \eqref{fullequations} receive contributions from $Q_{n}\big[\operatorname{T}_{E_{(s)}} W \big]$, $n \ge -2$, only. 

At this stage, we have all ingredients to explicitly compute the two traces \eqref{fullequations}. In order to write
the result in a compact form, we specify the endomorphisms as 
\be\label{endos}
E_{(2)} = \alpha \bar{R} \, , \qquad E_{(0)} = \beta \bar{R} \, . 
\ee
Denoting the derivative of $f_k(\bar{R})$ with respect to $\bar{R}$ by a prime and its $t$-derivative with a dot, the result for the traces then reads
\be\label{frtraces}
\begin{split}
 T_{(0)} = & \, \frac{k^6}{16 \pi^2} \, \frac{384 \pi^2}{\Rb^2} \, \frac{ c_1 f' + c_2 \, k^2 f'' + c_3  \dot{f}' + c_4 \, k^2 \dot{f}''}{
					\left(3 k^2-(3 \beta+1) \bar{R} \right)^2 f'' + \left(3 k^2-(3 \beta +2) \bar{R} \right) f' +2 f
				} \, , \\
  T_{(2)} = & \, \frac{k^6}{16 \pi^2} \, \frac{384 \pi^2}{\Rb^2} \, 
	\frac{
		\tilde{c}_1 f' + \tilde{c}_2 \,  \dot{f}'
	}{
		\left(3 k^2-(3 \alpha +1) \Rb \right) f'+3 f 
	} \, . 				
\end{split}
\ee
The coefficients are conveniently be expressed in terms of the dimensionless curvature $r \equiv \Rb k^{-2}$
and read
\spliteq{\label{cdef}
c_1 & = 3 + (1-6 \beta ) r  +\tfrac{1080 \beta ^2-360 \beta +29}{360} \, r^2 \, , \\[1.2ex]
c_2 &=
18 - 54 \beta r +\tfrac{3240 \beta ^2-91}{60} \, r^2 -\tfrac{3240 \beta ^3-273 \beta +29}{180} \, r^3 \, , \\[1.2ex]
c_3 &=
	\tfrac{1}{2} +\tfrac{1-6 \beta}{4} \, r +\tfrac{1080 \beta ^2-360 \beta +29}{720} \, r^2
%\\	&\qquad 
-\tfrac{45360 \beta ^3-22680 \beta ^2+3654 \beta -185}{90720}  \, r^3 \, , \\[1.2ex]
	c_4 &=
	\tfrac{9}{4} - 9 \, \beta \,  r +\tfrac{3240 \beta ^2-91}{240} \, r^2 -\tfrac{3240 \beta ^3-273 \beta +29}{360} \, r^3 +
\\
&\qquad \hspace{160pt}+\tfrac{90720 \beta ^4-15288
   \beta ^2+3248 \beta -181}{40320} \, r^4 \, . 
}
and 
\spliteq{\label{ctdef}
\tilde{c}_1 &=
	15  -  (30 \alpha +5) \, r + \tfrac{1080 \alpha ^2+360 \alpha -1 }{72} \, r^2 ,
\\
\tilde{c}_2 &=
	\tfrac{5}{2}-\tfrac{30 \alpha + 5}{4} \, r +\tfrac{1080 \alpha ^2+360 \alpha -1}{144} \, r^2
 -\tfrac{45360 \alpha ^3+22680 \alpha ^2-126 \alpha - 311}{18144} \, r^3 \, . 
}

The PDE governing the scale-dependence of $f_k(R)$ is then obtained by substituting
eqs.\ \eqref{frrhs} and \eqref{frtraces} into \eqref{eq:flowEH}. It is most conveniently
expressed in terms of the dimensionless quantities
\eq{
r \equiv \Rb \, k^{-2},
\qquad
f_k(R) \equiv k^4 \varphi_k \left( \Rb k^{-2} \right).
}
In terms of these, the flow becomes
\be\label{dimlesspde}
\begin{split}
	32 \pi^2  \left( \dot{ \varphi} + 4 \varphi - 2 r \varphi' \right) & \, = 
\frac{(\tilde{c}_1 + 2 \tilde{c}_2) \, \varphi' +  \tilde{c}_2 \, (\dot{\varphi}' - 2 r \varphi'') }{\left( 3-(3\alpha+1)r \right) \, \varphi' + 3 \, \varphi}  \\   
 & \,  
+ \frac{d_1\, \varphi' +  d_2\, \varphi'' + c_3\, \dot{\varphi}' + c_4 \, \left(\dot{\varphi}'' - 2 r \varphi''' \right)}{\left(3-(3\beta+1)r\right)^2 \varphi'' + \left( 3 - (3 \beta+2) r \right) \varphi' + 2 \varphi} \, ,
\end{split}
\ee
with $d_1 = c_1 + 2 c_3$ and $d_2 =c_2 -2rc_3$.
Again, it is implicitly understood that $\varphi \equiv \varphi_k(r)$ and primes denote derivatives with respect to $r$ while the dots are derivatives with respect to 
the renormalization group time $t = \ln(k)$. The polynomials $c_i$ and $\tilde{c}_i$ given in eqs.\ \eqref{cdef} and \eqref{ctdef}, respectively. Since the flow equation
has been derived on a spherical background with $\bar{R} > 0$, it is valid for $r \ge 0$. Eq.\ \eqref{dimlesspde} constitutes the central result of this section. 
%-------------------------------------------------------------------------

%--------------------------------------------
\section{The $f(R)$-truncation: properties of the RG flow}
\label{Sect.4}
%--------------------------------------------

We now proceed by analyzing the properties of the PDE \eqref{dimlesspde}, governing the scale-dependence 
of the dimensionless function $\varphi_k(r)$ by
%. We start by discussing the singularity structure 
%of the PDE and its consequences for the existence of global solutions in Sect.\ \ref{Sect:5.1}.
%Subsequently, we 
projecting the PDE onto the subspace spanned by polynomials of the curvature
scalar and explore the resulting fixed point structure. In this way, we identify
the NGFP generalizing the Einstein-Hilbert analysis (cf.\ Tab.\ \ref{tab:EHfp}). The properties of the NGFP
then depend on the endomorphisms introduced in the regularization procedure.
This dependence is studied in detail and minimized following the principle of minimum sensitivity (PMS),
``optimizing'' the value of the critical exponents found within a given truncation.

%---------------------------------------------------------------------------------------------
%\subsection{The NGFP and its properties}
%---------------------------------------------------------------------------------------------
In order to get a more profound picture of the fixed point structure entailed by the PDE
\eqref{dimlesspde}, we resort to a polynomial ansatz for $\varphi_k(r)$,
including powers of the dimensionless curvature up to $r^N$
\eq{\label{anspoly}
&\varphi_k (r)  = \sum_{m = 0}^{N} g_{m} \, r^m,
&\del_t\varphi_k (r) = \sum_{m = 0}^{N} \beta_{g_m} \, r^m \, . 
}
Here $g_{m}$ are the $k$-dependent running couplings and $\beta_{g_m} \equiv \del_t g_m$ denotes their beta functions.
Substituting \eqref{anspoly} into \eqref{dimlesspde} and expanding the result in powers of $r$ up to order $N$ yields a system of 
 $N+1$ algebraic equations which can be solved for the beta functions 
\be\label{betafctsext}
\beta_{g_m} = \beta_{g_m}(\{ g_j \},\alpha,\beta) \, .
\ee
The beta functions depend on the couplings $g_m$ and, owed to the inclusion of the endomorphisms
in the regulators, have a parametric dependence on $\alpha,\beta$.
At a fixed point of the RG flow $g^*_m$, the beta functions \eqref{betafctsext} vanish.
Owed to the parametric dependence on $\alpha,\beta$ the position of such a fixed point
$g^*_m(\alpha,\beta)$ and its stability coefficients $\theta_m(\alpha,\beta)$ (defined
as minus the eigenvalues of the stability matrix \eqref{defstab}) depend on the endomorphisms.

We now extend our analysis of the NGFP identified in Sect.\ \ref{sect:EH}
to the polynomial truncations \eqref{anspoly}.
This extension has to be carried out with care, since new, unphysical fixed points appear,
when increasing the order $N$ of the expansion.
Thus, it is necessary to identify the correct NGFP at low order ($N=1$) and increase 
the dimension of the truncation step by step. 
By comparing the values of the fixed points at the order $N$ with those determined at the previous order $N-1$, it is possible to 
trace the NGFP through the system \eqref{anspoly}.

In order to make contact with previous works and the Einstein-Hilbert truncation of  Sect.\ \ref{sect:EH}, we first
study the system \eqref{betafctsext} for vanishing endomorphisms $(\alpha=\beta=0)$.
The position of the NGFP for the orders $N=1$ to $N=6$ are summarized in the first block of Tab.\ \ref{tab:NGFPzeroendo}.
\begin{table}[t!]
\begin{center}
\begin{tabular}{|c|c|c|c|c|c|c|c|}
\hline
$N$ & $g_0^*$   & $g_1^*$     & $g_2^*$   &   $g_3^*$  & $g_4^*$    & $g_5^*$    & $g_6^*$ \\ \hline \hline
$1$ & $0.0103$ & $-0.0255$  &           &            &            &            & \\
$2$ & $0.0105$ & $-0.0225$  & $0.0025$ &            &            &            & \\
$3$ & $0.0105$ & $-0.0242$  & $0.0022$ & $-0.0086$ &            &            & \\
$4$ & $0.0102$ & $-0.0246$  & $0.0020$ & $-0.0095$ & $-0.0085$ &            & \\
$5$ & $0.0102$ & $-0.0247$  & $0.0020$ & $-0.0083$ & $-0.0083$ & $-0.0053$ & \\
$6$ & $0.0102$ & $-0.0247$  & $0.0020$ & $-0.0083$ & $-0.0082$ & $-0.0053$ & $-0.0002$ \\ \hline \hline
$1$ & $0.0110$ & $-0.0248$  &           &            &            &            & \\
$2$ & $0.0116$ & $-0.0235$  & $0.0026$ &            &            &            & \\
$3$ & $0.0112$ & $-0.0237$  & $0.0024$ & $-0.0145$ &            &            & \\
$4$ & $0.0111$ & $-0.0238$  & $0.0024$ & $-0.0146$ & $-0.0073$ &            & \\
$5$ & $0.0111$ & $-0.0237$  & $0.0024$ & $-0.0141$ & $-0.0068$ & $-0.0046$ & \\
$6$ & $0.0111$ & $-0.0237$  & $0.0024$ & $-0.0141$ & $-0.0070$ & $-0.0048$ & $0.0013$ \\ \hline \hline
$1$ & $0.0170$ & $-0.0250$  &           &            &            &            & \\
$2$ & $0.0117$ & $-0.0237$  & $0.0026$ &            &            &            & \\
$3$ & $0.0110$ & $-0.0238$  & $0.0023$ & $-0.0121$ &            &            & \\
$4$ & $0.0108$ & $-0.0240$  & $0.0023$ & $-0.0125$ & $-0.0077$ &            & \\
$5$ & $0.0108$ & $-0.0240$  & $0.0023$ & $-0.0119$ & $-0.0074$ & $-0.0046$ & \\
$6$ & $0.0108$ & $-0.0240$  & $0.0023$ & $-0.0119$ & $-0.0078$ & $-0.0047$ & $0.0016$ \\ \hline 
\end{tabular}
\caption{\label{tab:NGFPzeroendo}Position of the NGFP in the polynomial expansion for the equation with $\alpha=0$ and $\beta=0$ (upper box),
the equal lowest eigenvalue condition with $\beta = 1/6, \alpha = -1/2$ (middle box), 
and for the choice of endomorphisms \eqref{pmsvals} favored by the principle of minimum sensitivity (lower box).}
\end{center}
\end{table}
Notably, the NGFP exists for all values $N$ and its position converges rapidly when increasing the size of
the truncation $N$. The critical exponents of the NGFP obtained by evaluating \eqref{defstab}
are given in the first block Tab.\ \ref{tab:stabilityzeroendo}.
\begin{table}[t!]
\begin{center}
\begin{tabular}{|c|c|c|c|c|c|c|c|}
\hline
\quad $N$ \quad & \quad $\theta_{0}$ \quad & \quad $\theta_{1}$ \quad & \quad $\theta_{2}$ \quad & \quad $\theta_{3}$ \quad & \quad $\theta_{4}$ \quad & \quad $\theta_{5}$ \quad & \quad $\theta_{7}$ \quad \\ \hline \hline
$1$ & \multicolumn{2}{c|}{$2.93 \pm 2.97 \i$}  &        &         &           &           & \\
$2$ & \multicolumn{2}{c|}{$2.69 \pm 4.61 \i$}  & $8.72$ &         &           &           & \\
$3$ & \multicolumn{2}{c|}{$3.24 \pm 3.10 \i$}  & $1.79$ & $-8.09$ &           &           & \\
$4$ & \multicolumn{2}{c|}{$3.43 \pm 3.14 \i$}  & $1.53$ & \multicolumn{2}{c|}{$-6.45 \pm 2.92 \i$}  &           & \\
$5$ & \multicolumn{2}{c|}{$3.41 \pm 3.33 \i$}  & $1.55$ & \multicolumn{2}{c|}{$-4.03 \pm 8.12 \i$}  & $-5.03$   & \\
$6$ & \multicolumn{2}{c|}{$3.08 \pm 3.17 \i$}  & $1.52$ & \multicolumn{2}{c|}{$-2.78 \pm 11.21\i$}  & $-4.86$   & $-10.84$ \\ \hline \hline
$1$ & \multicolumn{2}{c|}{$3.06 \pm 3.73 \i$}  &        &         &           &           & \\
$2$ & \multicolumn{2}{c|}{$3.01 \pm 6.87 \i$}  & $4.86$ &         &           &           & \\
$3$ & \multicolumn{2}{c|}{$3.42 \pm 4.07 \i$}  & $1.71$ & $-10.08$ &           &           & \\
$4$ & \multicolumn{2}{c|}{$3.68 \pm 4.22 \i$}  & $1.66$ & \multicolumn{2}{c|}{$-9.60 \pm 3.13 \i$}  &           & \\
$5$ & \multicolumn{2}{c|}{$3.59 \pm 4.30 \i$}  & $1.68$ &  \multicolumn{2}{c|}{$-6.36 \pm 14.43 \i$}  &  $-5.70$ & \\
$6$ & \multicolumn{2}{c|}{$3.51 \pm 4.25 \i$}  & $1.80$ & \multicolumn{2}{c|}{$-3.76 \pm 16.33\i$}  & $-5.61$   & $-13.73$ \\ \hline \hline
$1$ & \multicolumn{2}{c|}{$3.00 \pm 3.42 \i$}  &        &         &           &           & \\
$2$ & \multicolumn{2}{c|}{$2.47 \pm 5.93 \i$}  & $4.74$ &         &           &           & \\
$3$ & \multicolumn{2}{c|}{$3.35 \pm 3.73 \i$}  & $1.67$ & $-9.42$ &           &           & \\
$4$ & \multicolumn{2}{c|}{$3.59 \pm 3.83 \i$}  & $1.61$ & \multicolumn{2}{c|}{$-7.50 \pm 4.10  \i$} &         & \\
$5$ & \multicolumn{2}{c|}{$3.54 \pm 3.96 \i$}  & $1.63$ & \multicolumn{2}{c|}{$-4.58 \pm 10.95 \i$} & $-5.78$ & \\
$6$ & \multicolumn{2}{c|}{$3.40 \pm 3.90 \i$}  & $1.61$ & \multicolumn{2}{c|}{$-2.40 \pm 11.85 \i$} & $-5.67$ & $-12.14$ \\  \hline 
\end{tabular}
\caption{\label{tab:stabilityzeroendo} Stability coefficients associated with the NGFP identified in Tab.\ \ref{tab:NGFPzeroendo} with $\alpha=0$ and $\beta=0$ (upper box),
the equal lowest eigenvalue condition with $\beta = 1/6, \alpha = -1/2$ (middle box), 
and for the choice of endomorphisms \eqref{pmsvals} favored by the principle of minimum sensitivity (lower box).}
\end{center}
\end{table}
Again we observe a rapid convergence of the critical exponents with increasing value $N$. Moreover, 
extending the system beyond $N>2$ does not give rise to further relevant deformations, characterized
by Re$\theta > 0$: the number of relevant deformations stabilizes at three. This is a strong
indication that classical power counting still constitutes a good ordering principle for
the relevance of operators at the NGFP. Operators which are power-counting irrelevant
at the classical level do not correspond to relevant deformations at the NGFP.
All these findings are in complete agreement with earlier studies 
based on non-geometric flow equations \cite{Machado:2007ea, Codello:2008vh}. 
It is quite remarkable, however, that for $N=2$ the critical exponent $\theta_2$ is much closer to the
values found for $N > 2$, indicating that the geometric flows studied here are less sensitive to such outliers.

Based on the confidence obtained from the flow equation with $\alpha = \beta = 0$,
we now carry out a systematic analysis on the influence of the endomorphisms.    
Since $\alpha$ and $\beta$ have been introduced via the cutoff functions, the variation of these parameters corresponds
to a change in the regularization procedure. Naturally, 
potential observables should not depend too strongly on the regularization scheme.
Therefore the dependence of the beta functions on $\alpha$ and $\beta$
may be exploited to optimize the value of the critical exponents
obtained within a given truncation by 
minimizing their dependence on $\alpha$ and $\beta$.
This logic follows the principle of minimal sensitivity (PMS).

In practice, we implement this PMS as follows. 
First we choose the set of observables whose regulator-dependence should be minimized.
We pick the real part of the relevant stability coefficients, Re$\theta_1$ and $\theta_2$
as well as the ``universal product'' \cite{Lauscher:2001ya}
\be
\tau_* \equiv g_* \, \lambda_* = \frac{g_0^*}{32 \, \pi \, (g_1^*)^2} \, .
\ee
Subsequently, we expand $\varphi_k(r)$ to the order $N=3$ and compute these
quantities as functions of $\alpha$ and $\beta$.
In the first analysis we confine ourselves to a one dimensional subspace 
of regulators obeying the condition of equal lowest eigenvalues \eqref{elev} by setting
$\alpha = \beta -\tfrac{2}{3}$.
\begin{figure}[t!]
\begin{center}	
\includegraphics[width=0.47\textwidth]{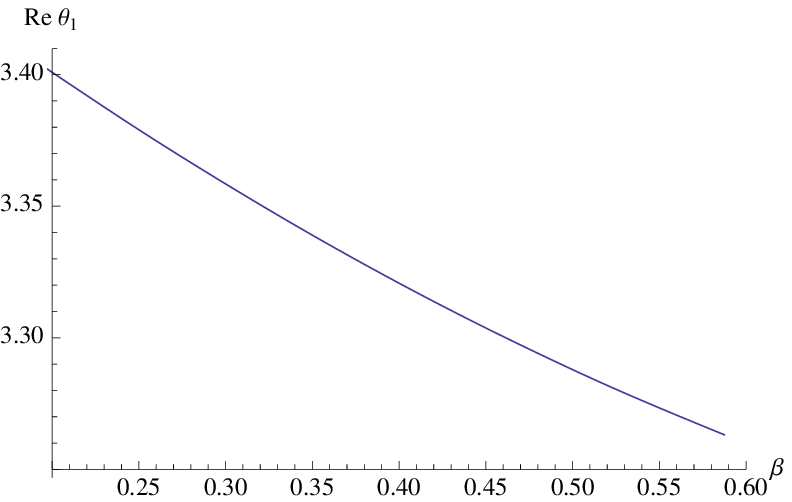}
\includegraphics[width=0.47\textwidth]{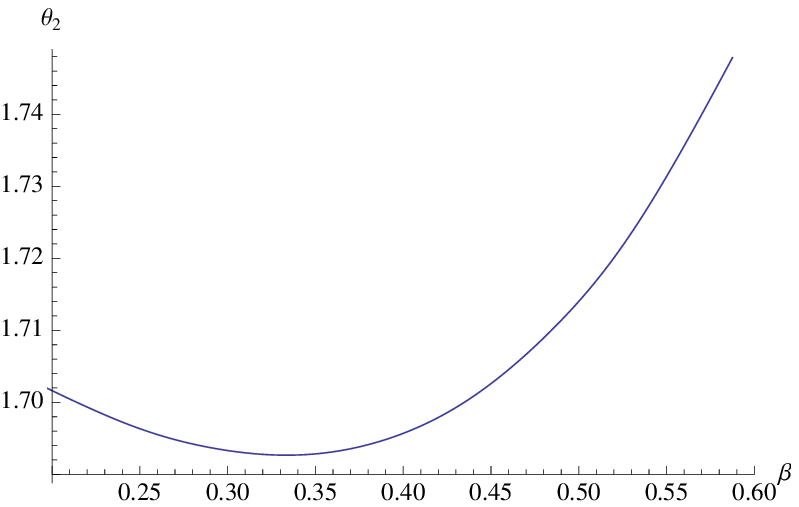} \\[3ex]
\includegraphics[width=0.47\textwidth]{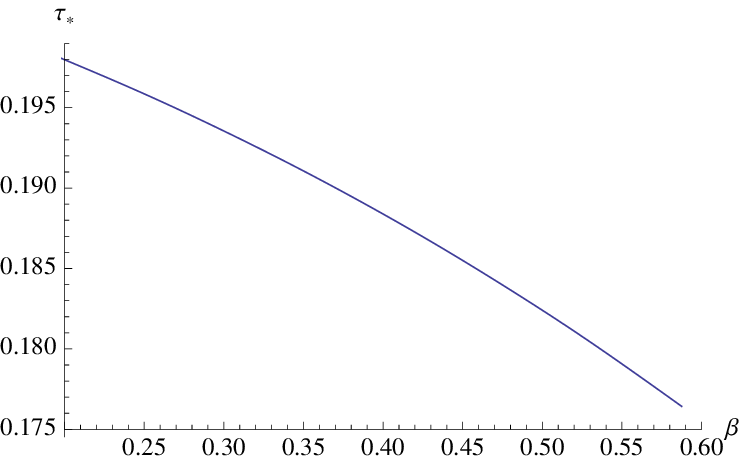}
\caption{\label{fig:theta123}
	Stability coefficients Re $\theta_1$, $\theta_2$ and $\tau_*$ as a function of the endomorphisms $\alpha$ and $\beta$
along the line of equal lowest eigenvalues $\alpha = \beta - \frac{2}{3}$.
}
\end{center}
\end{figure}
The values of $\operatorname{Re}\theta_{1}$, $\theta_2$ and $\tau_*$ along this line are displayed in Fig.\ \ref{fig:theta123}.
While $\operatorname{Re}\theta_{1}$ and $\tau_*$ show a monotonic behavior,  $\theta_2(\beta)$ shows a minimum at 
\be\label{pmsvals}
\beta \approx 0.332683 \, , \quad \alpha \approx -0.333984   \, . 
\ee
The value of $\tau_*$ decreases mildly with growing $\beta$. The range of values $0.175 \ge \tau_* \ge 0.2$ 
found in the linear-geometric case is similar to the values for $\tau_*$ found within 
 previous non-geometric studies \cite{Lauscher:2001ya,Codello:2008vh}, adding further confidence
to the linear-geometric approximation.

Now that we have identified a preferred choice for the endomorphisms, we repeat our fixed point
study for the distinguished choice \eqref{pmsvals}. 
The position of the NGFP $g_m^*$ and its critical exponents $\theta_m$ are shown in the lower boxes of Tab.\ \ref{tab:NGFPzeroendo} and Tab.\ \ref{tab:stabilityzeroendo},
respectively. Comparing the upper and lower parts of the tables establishes that the NGFP and its properties display
a weak dependence on the endomorphisms only: the optimized values differ very little from the $\alpha=\beta=0$ case.
Thus the polynomial $f(R)$-computation shows a strong robustness
with respect to varying the regulator functions.
Moreover, the PMS choice \eqref{pmsvals} significantly reduces the outlier $\theta_2$ found for $N=2$ (reported to be $8.4 \le \theta_2 \le 28.8$ in \cite{Lauscher:2002sq}) is further 
decreased, so that its value is only one fifth of the one found in non-geometric computations \cite{Machado:2007ea, Codello:2008vh}.
This analysis indicates that the Type I regularization procedure corresponding to vanishing endomorphisms $(\alpha=\beta=0)$,
is actually not favored by PMS: including a non-trivial endomorphism reduces the dependence of observables on the regulator
and improves the convergence properties of the polynomial truncations.

%--------------------------------------------
\section{Summary and Outlook}
\label{Sect.6}
%--------------------------------------------
We proposed a novel renormalization group equation for Quantum Einstein Gravity (QEG) of geometric type.
The construction involves a reference background, over which fluctuations of the metric are integrated out 
following the ideas of the Wilsonian renormalization group (RG).
The geometric setting eliminates redundant gauge configurations through
an opportune choice of coordinates on the configuration space: 
the coordinates are adapted to the
fiber-bundle structure and the gauge degrees of freedom are removed
by restricting the integration measure to the base space.
Consequently, the flow equation is formulated in terms of fluctuation fields, 
which are invariant with respect to quantum gauge transformations
and transform covariantly with respect to background gauge transformations.
 The resulting flow is manifestly invariant under both symmetries
and all expectation values are gauge invariant by construction.
Both the original Wetterich-type flow equations for gravity \cite{Reuter:1996cp,Lauscher:2001ya}
and the geometric flow equation constructed in this paper
are exact functional renormalization group equations for QEG,
which allow for the exploration of the theory's RG flow away from the Gaussian fixed point.
The structure of the geometrical
flow equation is considerably simpler, however,
since there are no contributions from the
gauge degrees of freedom, ghost fields and
auxiliary fields encoding the Jacobians of the TT-decomposition.

 In order to be able to perform explicit computations, we chose a generic background metric
as the base point and decomposed the fluctuations via a transverse-traceless decomposition
with respect to this background. Subsequently, we approximated the exact geometric flow
by truncating the map between fluctuations and gauge-invariant fields at the linear level.
In this ``linear geometric approximation'', the gauge-invariant fields of the geometric construction coincide
with those of the transverse-traceless decomposition.
From a geometric viewpoint, this approximation neglects the contributions
coming from the Vilkovisky-De Witt connection of the geometric formalism,
that appears upon projecting the redundant configuration space of the fields onto the space of physically distinguished configurations.
The flow obtained in this way is driven by quantum fluctuations of the transverse-traceless and trace part of the fluctuation field and
 agrees with the functional renormalization group equation \cite{Reuter:1996cp} 
for a specific choice of regulator function and gauge.

In this work we have performed a first set of non-trivial tests of the Asymptotic Safety conjecture based on the linearized geometric approximation.
At the level of the four-dimensional Einstein-Hilbert truncation the standard phase diagram of QEG \cite{Reuter:2001ag} is recovered
and the critical exponents associated with the non-Gaussian fixed point (NGFP) controlling the UV-behavior of the theory,
turn out quantitatively similar.  
Our results can also be analytically continued in the dimensionality of the system,
allowing to explore all dimensionalities and follow the NGFP up to the upper critical dimension of gravity.
Notably, the geometric beta functions possess a discontinuity when approaching the lower critical dimension, $d=2$,
which is related to the fact that there are no transverse-traceless tensors on a two-dimensional sphere.

One of the key advantages of the geometric flow equation becomes apparent
at the level of functional truncation including an infinite number of coupling constants.
Constructing the partial differential equation (PDE) that governs the RG flow of 
the function $f_k(R)$, 
the absence of gauge degrees of freedom, ghosts, and auxiliary fields
implies that the singularity structure of the PDE 
is completely determined by the contributions of the (approximately)
gauge-invariant fields. As it will turn out \cite{Demmel:ip},
the resulting structure is precisely the one needed 
for obtaining isolated and globally well-defined fixed functionals.

%  Applying
% the singularity counting theorem \cite{Dietz:2012ic}, it is easy to see
% that this gives rise to the singularity structure expected for 
% a countable number of 
% The explicit construction of the corresponding fixed functionals
% will be carried out in a companion paper.

For the purpose of this work, however, we limited ourselves to a RG flow study
performing a polynomial expansion of $f_k(R)$ which incorporates
 higher powers of the Ricci scalar in the truncation ansatz.
 As a novel feature, our construction
 of the beta functions includes a non-trivial endomorphism
 in the regularization procedure. Systematically increasing the order of the polynomial
 reliably identifies a unique NGFP. 
Including an increasing number of power-counting irrelevant operators, we establish that the fixed point comes with three relevant deformations.
Its position and critical exponents are in good agreement
with earlier studies based on the non-geometric constructions \cite{Codello:2007bd,Machado:2007ea,Codello:2008vh,Falls:2014tra}.
The finite (non-increasing) number of relevant deformations in the geometric flow equation is another signal underpinning 
the predictivity of Asymptotic Safety. Moreover, the coherence with earlier findings also corroborates the validity of our construction.

Notably, the properties of the NGFP show very little sensitivity towards the change of the
regularization procedure. The parametric dependence of the beta functions
on the endomorphisms allows to apply the principle of minimal sensitivity to the stability coefficients of the NGFP.
This principle allows to determine what set of parameters is numerically preferred
by minimizing the highest real critical exponent of the NGFP,
meaning that the system converges more rapidly to its critical point in a statistical mechanical sense \cite{Canet:2003qd}.
The results become thus more reliable because it can be argued that there is less sensitivity on the operators
that are neglected from the projection procedure.
Interestingly, this work shows that a regulator that utilizes
operators with non-zero endomorphisms (Type II cutoffs in the language of \cite{Codello:2008vh}) is preferred
over a regularization procedure based on Laplacian operators only (Type I cutoff).

Clearly, the geometrical flow equation may play an important 
role when computing vacuum expectation values of gauge-invariant
combinations of the fluctuation fields.
Moreover, keeping track of the fluctuation fields (so-called bi-metric computations) may be carried out more economically,
especially when exploiting the Ward identities of the construction \cite{Donkin:2012ud,Bridle:2013sra}.
Thus the geometric flow equation may be helpful in
understanding the actual driving elements of the gravitational RG flow,
which might have been previously overshadowed by gauge-effects.
In order to perform such computations,
it is clear that the natural generalization of the method
adopted in this paper is to take into account the non-vanishing 
curvature of the fiber-bundle structure underlying the gravitational configuration space.
This requires the knowledge of the map \eqref{map1} at least to quadratic order, while 
the computation of gauge-invariant vacuum expectation values of the fluctuation fields will require
going beyond the quadratic approximation.  We hope to come back to this point
in the future. 
%--------------------------------------------
\section*{Acknowledgments}
%--------------------------------------------
We thank A.\ Nink and M.\ Reuter for many helpful discussions.
The research of F.~S.\ and O.~Z.\ has been supported by the Deutsche Forschungsgemeinschaft (DFG)
within the Emmy-Noether program (Grant SA/1975 1-1).

%--------------------------------------------
\begin{appendix}
%--------------------------------------------
%--------------------------------------------
\section{The heat-kernel on the $d$-sphere $S^d$}
\label{app:A}
%--------------------------------------------
The explicit computation of the gravitational beta functions
requires the evaluation of the functional traces appearing
on the r.h.s.\ of the flow equation \eqref{geoflow}. These
calculations are conveniently done by applying results for the 
heat-kernel on a $d$-sphere and we collect the relevant formulas in this appendix.

%--------------------------------------------
\subsection{Heat-kernel on $S^d$: early time expansion}
\label{app:A2}
%--------------------------------------------
Throughout this paper we chose the background metric $\gb_{\mu\nu}$
to be the one of the $d$-dimensional sphere. This implies
that the curvature tensors constructed from $\gb_{\mu\nu}$
satisfy
\be\label{background}
\bar{R}_{\mu\nu\rho\sigma} = \frac{\bar{R}}{d(d-1)} \left( \gb_{\mu\rho} \, \gb_{\nu\sigma} - \gb_{\mu\sigma} \, \gb_{\nu\rho} \right) \, , \qquad \bar{R}_{\mu\nu} = \frac{1}{d} \, \bar{g}_{\mu\nu} \, \bar{R} \, ,  
\ee
and are covariantly constant. Moreover, the volume of the $d$-sphere is related to the Ricci scalar by
\eq{\label{Sdvol}
{\rm Vol}_{S^d} \equiv \int_{S^d}\diff^d x \sqrt{\gb} = \frac{\Gamma(d/2)}{\Gamma(d)} \left( \frac{4 \,\pi \, d(d-1)}{\bar{R}} \right)^{d/2}.
}

The fact that the background curvatures are covariantly constant allows to relate a function $W(x)$ depending
on the covariant background Laplacian $\Delta \equiv - \gb^{\mu\nu} \bar{D}_\mu \bar{D}_\nu$
 to the heat-kernel of $\Delta$ in a rather simple way
\be\label{meltraf0}
	\Tr_{(s)}\left[ W(\Delta) \right] = \int_0^\infty\diff \sigma\; \widetilde{W}(\sigma) \Tr_{(s)} \left[ \E^{-\sigma\Delta} \right] \, . 
\ee
Here $\widetilde{W}$ is the inverse Laplace transform of $W$
and the subscript $(s)$ on the traces indicates whether $\Delta$ acts 
on symmetric transverse traceless tensors ($s=2$), transverse vectors ($s=1$), or scalars $(s=0)$.
The heat trace admits an early-time expansion 
 \eq{
\label{eq:asymptotic}
\Tr_{(s)}\left[ \E^{-\sigma \Delta} \right] \simeq \frac{1}{(4\pi \sigma)^{d/2}} \int_{S^d}\diff^d x \sqrt{\gb}\;  \sum_{n\geq 0} \, b^{(s)}_n \, \sigma^n \,  \bar{R}^n \, .
}
The expansion coefficients $b^{(s)}_n$ depend on the spin of the field. Following
ref.\ \cite{Lauscher:2001ya} they can be obtained from the early-time expansion
of tensor fields without differential constraints. The first two coefficients
obtained in this way are 
\spliteq{
\label{eq:EHhkcoeff}
b^{(0)}_0 &= 1, \qquad &b^{(0)}_1 &= \frac{1}{6},
\\
b^{(2)}_0 &= \frac{(d-2)(d+1)}{2}, \qquad &b^{(2)}_1 &= \frac{(d+1)(d+2)(d-5 + 3 \delta_{d,2})}{12(d-1)}.
}

Substituting \eqref{eq:asymptotic} into \eqref{meltraf0}, the operator trace can be written as
\spliteq{\label{mastertrace}
 \Tr_{(s)} \left[ W(\Delta) \right]
&= 
\frac{1}{(4\pi)^{d/2}}\int\diff^d x \sqrt{\gb}\; \sum_{n\geq 0} Q_{d/2-n}[W]\, b^{(s)}_n\,  \bar{R}^n \, ,
}
with the $Q$-functionals defined as
\be\label{Qdef}
Q_n[W] \equiv
\int_0^\infty \diff \sigma\, \sigma^{-n} \, \widetilde{W}(\sigma)
\ee
For $n > 0$ this definition can be related to $W$ by a Mellin-transform
\be\label{Qeval1}
\begin{split}
Q_n[W] = & \, \frac{1}{\Gamma(n)} \, \int_0^\infty \diff z \, z^{n-1} \, W(z) \, , \qquad n > 0 \, , \\
Q_{-m}[W] = & \, (-1)^m \, W^{(m)}(0) \, , \qquad \qquad  \quad m \ge 0 \in \mathbb{N} \, , 
\end{split}
\ee 
with $W^{(m)}(z)$ denoting the $m$th derivative of $W$ with respect to the argument. Based on
these formulas, the derivation of the beta functions \eqref{betafcts} is rather straightforward.

%--------------------------------------------
\subsection{Heat-kernel on $S^4$: the asymptotic series}
\label{app:A1}
%--------------------------------------------
%
\begin{table}[t]
\centering
\renewcommand\arraystretch{1.5}
\begin{tabular}{|c|c|c|l|}
\hline
Spin $s$ & Eigenvalue $\lambda_l(d,s)$ & Multiplicity $M_l(d,s)$ & \\ \hline \hline
0    & $ \frac{l(l+d-1)}{d(d-1)}R $ & $\frac{(2l+d-1)(l+d-2)!}{l!(d-1)!}$  & $l = 0,1,\dots$ \\ \hline
2    & $\frac{l(l+d-1)-2}{d(d-1)}R$  & $\frac{(d-2) (d+1) (l-1) (d+l) (d+2 l-1) (d+l-3)!}{2 (d-1)! (l+1)!}$ & $l = 2,3,\dots$ \\ \hline
\end{tabular}
\caption{Eigenvalues $\lambda_l(d,s)$ and multiplicities $M_l(d,s)$ of the Laplacian operator $\Delta \equiv -D^2$ acting on fields with spin $s$ on the $d$-sphere \cite{Rubin:1983be,Rubin:1984tc}.}
\label{tab:spectrum}
\end{table}
The derivation of the PDE governing
the scale-dependence of $f_k(R)$ in Sect.\ \ref{Sect.2.1} requires
knowing the expansion coefficients $b_n^{(s)}$ for $s=0,2$ to higher order in $n$.
The coefficients can be found by relating the heat kernel to
the sum over eigenvalues $\lambda_l(d,s)$ of the operator $\Delta$
weighted by their multiplicity $ M_l(d,s)$
\eq{\label{evsum}
	\Tr_{(s)} \left[ \E^{-\sigma \Delta} \right] = \sum_l \, M_l(d,s) \, \E^{-\sigma \lambda_l(d,s)} \, .
}
For general $d$, $\lambda_l(d,s)$ and $M_l(d,s)$ have been computed in \cite{Rubin:1983be,Rubin:1984tc} and are listed in Tab.\ \ref{tab:spectrum}.
We stress that for $d$ even, \eqref{evsum} only constitutes an asymptotic series \cite{Avramidi:2000bm}. The first two terms in the expansion
are universal while the coefficients multiplying higher powers of $R$ may depend on the resummation scheme. We fix this freedom
by demanding that \eqref{evsum} reproduces the early-time expansion \eqref{eq:asymptotic} evaluated on the $d$-sphere.

The case where \eqref{evsum} reproduces
the early-time expansion of the heat kernel
evaluates the sum using the Euler-MacLaurin formula
\eq{
\label{eq:EulerMacLaurin}
\sum_{n=a}^b f(n)
\sim
\int_a^b f(x)\, \diff x + \frac{f(b)+f(a)}{2}
+ \sum_{k=1}^{\infty} \frac{B_{2k}}{(2k)!} \left( f^{(2k-1)}(b) - f^{(2k-1)}(a)\right) \, .
}
Here $B_{2k}$ denotes the Bernoulli numbers. For $d=4$ the functions $f^{(s)}(x)$
entering into \eqref{eq:EulerMacLaurin} are
\be
\begin{split}
f^{(0)}(x) = & \, \frac{1}{6} (x+1) (x+2) (2 x+3) \E^{-\frac{1}{12}   x (x+3)R \sigma} \, , \\
f^{(2)}(x) = & \, \frac{5}{6} (x-1) (x+4) (2 x+3) \E^{-\frac{1}{12}  (x (x+3)-2)R \sigma}\, , \\
\end{split}
\ee
with boundaries $a=0, b=\infty$ ($s=0$) and $a=2, b=\infty$ ($s=2$). The 
 integral parts in \eqref{eq:EulerMacLaurin} are then given by 
\be
\begin{split}
\int_0^\infty \diff x \, f^{(0)}(x)
= & \,
\frac{1}{(4\pi\sigma)^2}\int_{S^d}\diff^d x \sqrt{\gb}\; \left( 1+ \tfrac{1}{6}\sigma \bar{R} \right) \, , \\
\int_2^\infty \diff x \, f^{(2)}(x)
= & \, 
\frac{1}{(4\pi\sigma)^2}\int_{S^d}\diff^d x \sqrt{\gb}\;\, \left( 5 + \tfrac{5}{2} \sigma \bar{R}  \right) \, \E^{-\frac{2 \bar{R} \sigma }{3}}
\end{split}
\ee
where we have reinstalled the volume integral via the relation \eqref{Sdvol}.
In contrast to the scalar case, where the integral part determines the first two heat-kernel coefficients,
the integral for $s=2$ contributes to all orders in $\bar{R}$. The sums in \eqref{eq:EulerMacLaurin} start contributing at order $\bar{R}^2$ in the expansion
and the corresponding coefficients can be computed by evaluating them on a term by term basis truncating the infinite sufficiently high order.
This procedure yields the heat-kernel coefficients $b_n^{(s)}$ listed in Tab.\ \ref{tab:HKcoeffs}. 
\begin{table}[t]
	\centering
	\renewcommand\arraystretch{1.5}
	\begin{tabular}{|c | c| c| c |c| c| c|}
	\hline 
		s & $b_0^{(s)}$ & $b_1^{(s)}$ 	  	& $b_2^{(s)}$ 		& $b_3^{(s)}$		        &           $b_4^{(s)}$ 	 	&          $b_5^{(s)}$ \\ \hline \hline
		0 &  $1$ & $\frac{1}{6}$  	&  $\frac{29}{2160}$  	& $\frac{37}{54432}$    & $\frac{149}{6531840}$    	& $\frac{179}{431101440}$ \\ \hline
		2 &  $5$ & $-\frac{5}{6}$ 	& $-\frac{1}{432}$	& $\frac{311}{54432}$	& $\frac{109}{1306368}$		& $-\frac{317}{12317184}$ \\ \hline
	\end{tabular}
	\caption{Heat kernel coefficients appearing in the early-time expansion of \eqref{eq:asymptotic} on the four-sphere.}
	\label{tab:HKcoeffs}
\end{table}
These coefficients form the basis for constructing the PDE governing the scale-dependence of $f_k(R)$ in Sect.\ \ref{Sect.2.1}.
%--------------------------------------------
\end{appendix}
%--------------------------------------------

%--------------------------------------------
\end{document}